\documentclass[fleqn,usenatbib]{mnras}

\usepackage[T1]{fontenc}
\usepackage{ae,aecompl}

\usepackage{graphicx}	

\newcommand{\MSun}{\mbox{${\rm M}_\odot$}}
\newcommand{\pc}{\mbox{${\rm pc}$}}
\newcommand{\kms}{\mbox{${\rm km/s}$}}

\title[Young sub-clusters]{Dynamical evolution of stars and gas of young embedded stellar sub-clusters}

\author[A. Sills et al.]{Alison Sills$^{1}$\thanks{E-mail: asills@mcmaster.ca (AS)},
Steven Rieder$^{2}$,
Jennifer Scora$^{1}$,
Jessica McCloskey$^{1}$ and
Sarah Jaffa$^{3}$
\\
$^{1}$Department of Physics \& Astronomy, McMaster University, 1280 Main Street West, Hamilton ON, L8S 4M1, CANADA\\
$^{2}$RIKEN Advanced Institute for Computational Science, 7-1-26 Minatojima-minami-machi, Chuo-ku, Kobe, 650-0047, Hyogo, Japan\\
$^{3}$School of Physics and Astronomy, Cardiff University, The Parade, Cardiff, CF24 3AA, UK\\
}

\date{Accepted 2018 March 8. Received 2018 March 7; in original form 2017 November 17}

\pubyear{2018}

\begin{document}\label{firstpage}
\pagerange{\pageref{firstpage}--\pageref{lastpage}}
\maketitle

\begin{abstract}
We present simulations of the dynamical evolution of young embedded star clusters. Our initial conditions are directly derived from X-ray, infrared, and radio observations of local systems, and our models evolve both gas and stars simultaneously. Our regions begin with both clustered and extended distributions of stars, and a gas distribution which can include a filamentary structure in addition to gas surrounding the stellar subclusters. We find that the regions become spherical, monolithic, and smooth quite quickly, and that the dynamical evolution is dominated by the gravitational interactions between the stars. In the absence of stellar feedback, the gas moves gently out of the centre of our regions but does not have a significant impact on the motions of the stars at the earliest stages of cluster formation. Our models at later times are consistent with observations of similar regions in the local neighbourhood. We conclude that the evolution of young proto-star clusters is relatively insensitive to reasonable choices of initial conditions. Models with more realism, such as an initial population of binary and multiple stars and ongoing star formation, are the next step needed to confirm these findings.
\end{abstract}




\section{Introduction}

Canonical wisdom in the field of star formation is that ``all stars form in clusters''.
A more accurate summary of the paper used to support that statement  \citep{LadaLada2003} is that most Galactic star formation occurs in embedded protoclusters, many of which will not emerge from their molecular clouds as bound star clusters.
Certainly it is true that stars do not form in truly isolated regions.
Both observations of young star-forming regions \citep[e.g.][]{MYSTIXOverview2013} and theory and simulations of star formation \citep[e.g.][]{Offner2009} paint a picture of star formation as a turbulent, clumpy, stochastic process.
The crowded environment in which stars form must, to some extent, determine the properties of stars themselves -- the initial mass function, stellar multiplicity distributions, and probably their planetary properties as well.
At the same time, we expect that some of the protoclusters manage to remain as gravitational bound entities, and should form the open clusters we see today. Embedded clusters may also provide insights relevant to the earliest stages of the life of massive, dense globular clusters, especially now that we are learning that those clusters may have had more than one episode of star formation \citep{Gratton2012}. 

There are a number of as-yet unanswered questions about the processes through which embedded protoclusters evolve.
For example, most open clusters have little remaining gas and dust, and are well-described by a smooth, spherical stellar spatial distribution.
Younger embedded clusters, on the other hand, more often show clustered or filamentary substructure.
The Massive Young Stellar complexes study in Infrared and X-rays (MYStIX) \citep{MYSTIXOverview2013} identified over 31 000 stars in 20 nearby star-forming regions, and used mixture models to identify (sub)clusters of isothermal ellipsoids in each region \citep{Kuhn2014}.
These 20 regions contain 142 subclusters, and each also has a non-clustered, distributed population of stars.
A careful analysis allowed the MYStIX team to determine the total number of stars, the size, the ellipticity, and the approximate age of each subcluster \citep{Getman2014}.
These regions display a variety of morphologies.
Some are clearly filamentary, with a string of sub-clusters creating a linear structure on the sky.
Others are more centrally concentrated in one dominant cluster with a core-halo morphology, and still others are characterized as clumpy.
How and when do the regions lose their substructure?  In the MYStIX observations, there is no obvious age difference between the most substructured systems and the most spherical regions.
On the other hand, merging is the outcome of many simulations of clumpy or fractal self-gravitating star systems \citep{Proszkow2009, Fujii2012, Banerjee2015}.
Models suggest that the erasure of small-scale structure through merging can affect quantities that are used to trace the process of star formation, such as mass segregation \citep[e.g.][]{Parker2014}. 

Embedded clusters may also be affected by the remnants of their natal molecular cloud.
Recent observations of the Orion Nebula Cluster \citep{Stutz2017} demonstrate that the gas component of that cluster dominates the gravitational potential throughout the cluster except at the very centre.
In another of the MYStIX regions, DR21, the gas mass is at least comparable to the stellar mass \citep{Schneider2010}.
Within a few tens of Myr, however, star clusters and associations have a very low gas content.
It is expected that gas is removed by stellar feedback in the form of outflows, stellar winds, and eventually supernovae.
Some dynamical models of embedded clusters have studied this early gas loss by including an additional gravitational potential which is lost either instantaneously \citep{Proszkow2009} or by decreasing the gas mass with time \citep{Banerjee2015}.
Rapid loss of the gas potential can leave an unbound stellar association that disperses with time \citep{GoodwinBastian2006} in a stage commonly called `infant mortality', although recent simulations suggest that perhaps the effects of gas expulsion may be less dramatic than previously thought \citep{Parker2013}.
Very few simulations have followed the simultaneous evolution of gas and stars, and allowed both components to influence the motion of the other, although codes have been developed with this kind of project in mind \citep[e.g.\ the {\tt SEREN} code,][]{Hubber2013}.
The early work of \citet{Pelupessy2012} used the AMUSE code to model spherical clusters of 1000 stars suggested that the duration and method of gas removal could have a substantial effect on the eventual appearance and structure of the star cluster. 
Our theoretical understanding of embedded star clusters needs to involve both stars and gas, and should also be driven by our observations of such systems in the local Galaxy.

In this paper, we present simulations of embedded protoclusters, including the simultaneous evolution of both their stellar and gaseous components.
Our simulations explicitly use the observed sub-structure of embedded clusters, taken from the MYStIX work, to inform the initial conditions for our simulations.
We also use live gas particles, and calculate the gravitational interaction between the gas and the stars in these embedded clusters.
We are interested in understanding the timescales and processes that drive the evolution of these systems, with the intention of eventually being able to understand how bound stellar systems across mass scales are assembled and evolve. 

In the next section, we describe the methods we used to set up and evolve young star clusters based on observational initial conditions.
We discuss our results in section 3, and draw some conclusions as well as outlining future work in section 4. 

\section{Method}

We used {\tt AMUSE} \citep{AMUSE2009,AMUSE2013}, the Astrophysical Multipurpose Software Environment, to create our initial conditions as well as to evolve the gas and stars simultaneously.
AMUSE is essentially a python wrapper that connects different community codes that model many physical effects: gravitational dynamics of both large and small numbers of particles, hydrodynamics, radiative transfer, and stellar and binary evolution, as the user requires.
In these simulations, we couple a collisional N-body gravitational dynamics code with a hydrodynamical code, and allow the star and gas particles to influence each other gravitationally through a mechanism called Bridge \citep{Fujii2007}.
We create the initial conditions in the AMUSE format based on the observational constraints we have, and then we evolve each system for up to 10 Myr.  

\subsection{Initial Conditions}\label{sec:initial} 

\begin{figure}
	\includegraphics[width=\columnwidth]{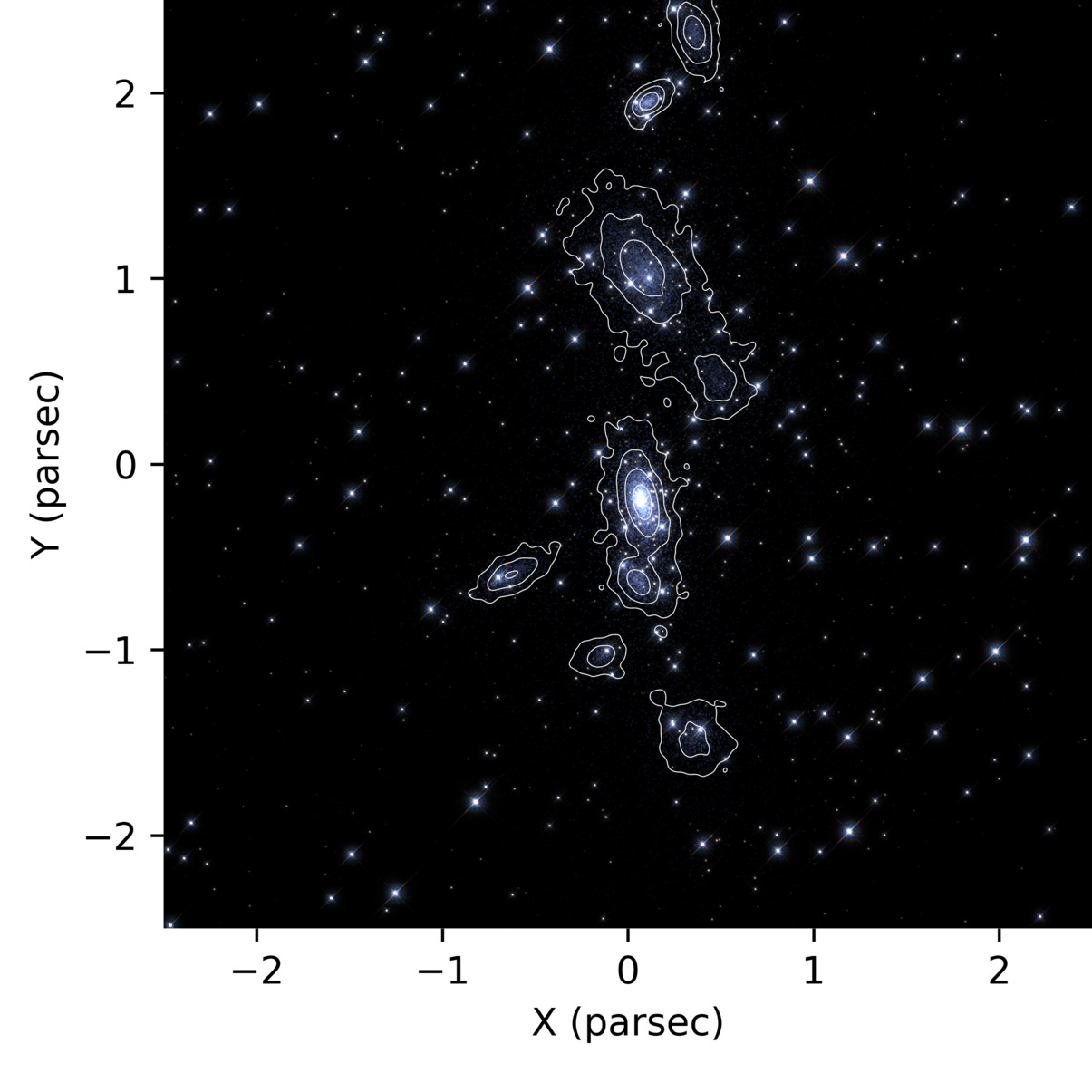}
  \caption[Initial conditions for DR21]{Initial conditions for DR21. 
    X and Y axes correspond to RA and Dec, respectively.
    Stars are shown as blueish-white objects; gas is included as diffuse emission, reflecting light from nearby bright stars.
    Gas is also indicated with contour lines. The lowest contour level corresponds to a surface density of $150~\MSun~\pc^{-2}$, and each subsequent line is a factor of $10^{0.5}$ higher.
    Figure created using {\tt FRESCO}\footnotemark.
    }\label{fig:DR21initial}
\end{figure}
\footnotetext{\url{https://github.com/rieder/FRESCO}}

Most previous simulations of young clusters have assumed either an initial spherical distribution of stars \citep{Proszkow2009}, a group of spherical sub-clumps \citep{Banerjee2015,Fujii2012} or a fractal distribution \citep{Parker2016} which more closely resembles the observations of young clusters.
In our case, we took our initial conditions directly from observations of young clusters where possible. 

The MYStIX observations, in \citet{Kuhn2014}, give the number of subclusters, the semi-major and semi-minor axes of each ellipse which best fits that subcluster as well as the position angle of each ellipse on the sky.
They also give the right ascension and declination of each star observed in the region.
The fit to the stellar density includes the sub-clusters as well as an `unclustered' component (essentially a uniform density background of stars).
There are also some stars in each region which are labelled as `unknown'.
These are stars that are not cleanly identified with either a sub-cluster or the uniform background, but still exist in the image and therefore are included in our simulations.

\subsubsection{Positions}\label{sec:initial-positions}
First, we used the stellar positions taken directly from the MYStIX observations.
We mapped the right ascension and declination of each star to x and y positions in parsec from the cluster centre, using the distances and cluster centres given in \citet{Kuhn2014}.
All stars in the MYStIX observations are used, including stars assigned to individual subclusters, and also the background of stars labelled as `unclustered' and `unknown'.
We assigned z (line of sight) positions of these stars by drawing randomly from a reasonable range of distances.
For the subclusters, we used the size of the minor axis of each subcluster as the z range for each star.
The z range for the background stars, $z_{\max}$ is set to a value which encompasses the maximum extent of these stars in the xy plane. 

MYStIX only measures positions for stars more massive than approximately $0.83~\MSun$. Therefore, we added a population of low-mass stars ($0.08$ to $0.83~\MSun$).
For the sub-clusters, the total number of stars was taken from \citet{Kuhn2015}, where they estimated the unseen stars by populating an initial mass function (IMF) normalized by the number of observed stars.
We create a \citet{Plummer1911} sphere, with a core radius given by the harmonic mean of the observed major and minor axes of the fitted ellipses. For a Plummer sphere with a scale radius $a$, the core radius is $\approx 0.64a$, the half-mass radius is $\approx 1.3a$, and the virial radius (used by AMUSE to set up the distribution) is $16a/(3\pi)$. We then multiply the position of each star along one axis by the necessary factor so that the Plummer sphere becomes the appropriate observed ellipse.
We also use the observed position angle of each sub-cluster to get the correct relative orientation of all sub-clusters in the plane of the sky.
For most of our simulations, we assumed that each subcluster lies along the z=0 plane, but we also varied this parameter to allow the subclusters to be scattered about the z=0 plane by a random distance that was less than our parameter $Z_{\rm offset}$. 
We also add a uniform density sphere of low-mass stars around the entire region, where again the total number of unclustered stars of all masses is calculated by populating an initial mass function down to $0.08~\MSun$, normalized by the number of higher-mass stars. Given the mass range of the observed stars ($0.83$ to $10~\MSun$) and our assumed IMF slope, we add approximately 2 low-mass stars for each observed star.
The size of this sphere is the same as the size we used to determine the z positions of the higher-mass unclustered stars. 

\begin{figure}
	\includegraphics[width=\columnwidth]{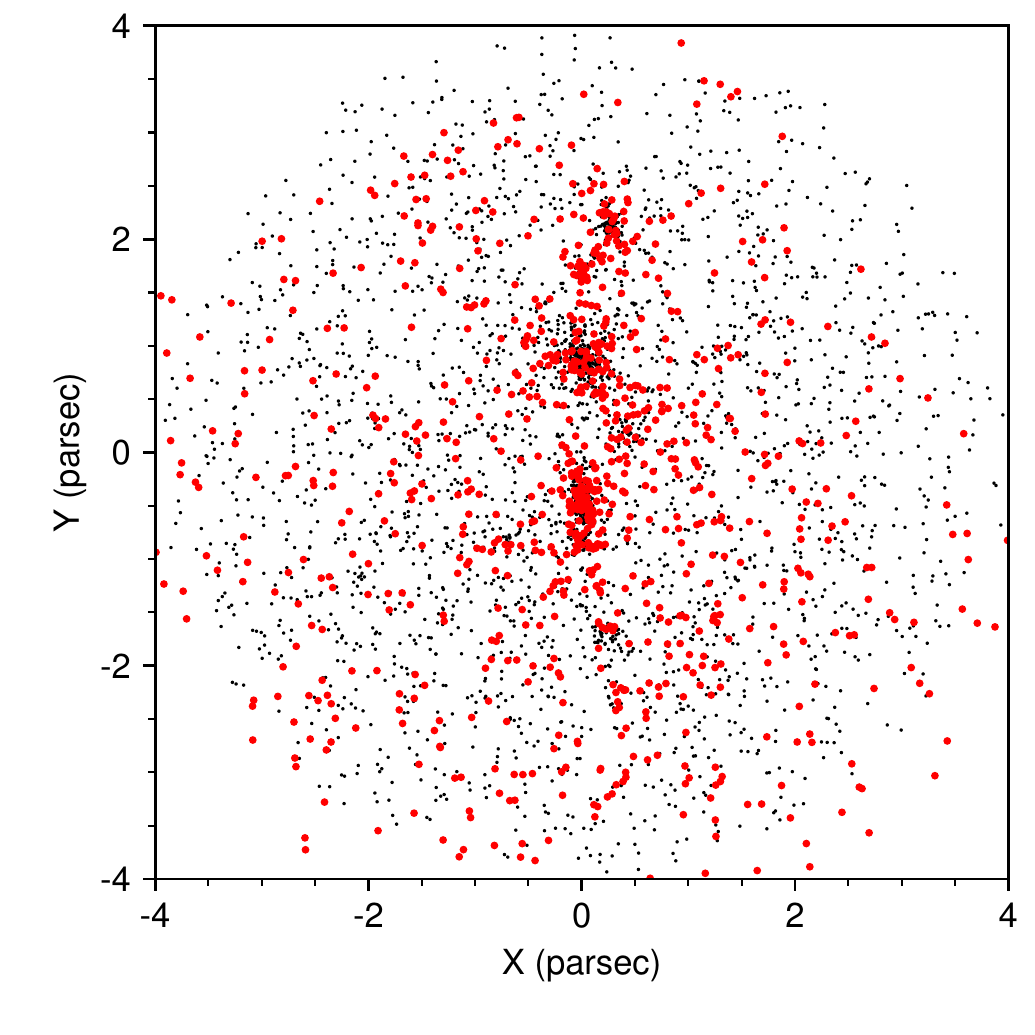}
  \caption{Initial stellar positions for the DR21 simulations. Observed MYStIX stars are shown as red circles, and the inferred population of low-mass stars are shown as small black points.
  }\label{fig:StarPositionsAll}
\end{figure}

\begin{figure}
	\includegraphics[width=\columnwidth]{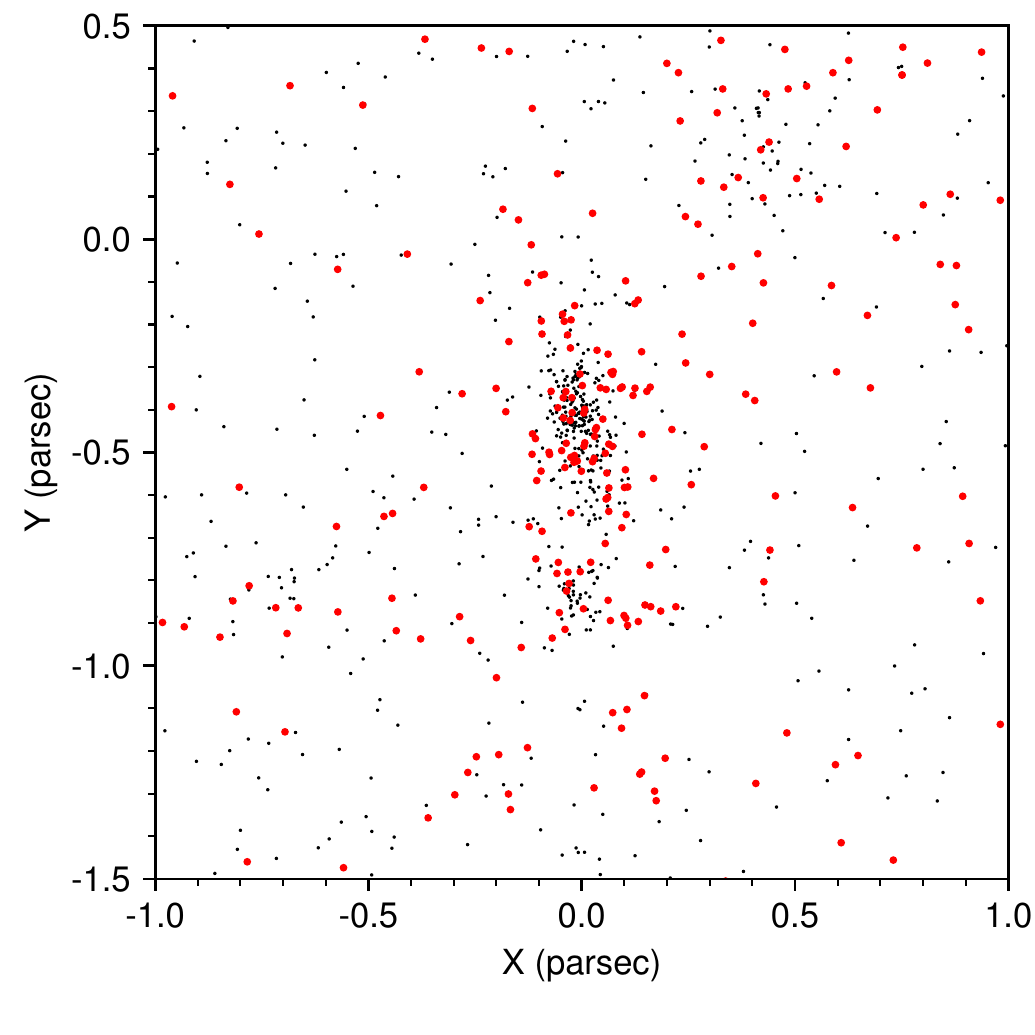}
  \caption{Initial stellar positions for one sub-cluster in the DR21 simulations. Observed MYStIX stars are shown as red circles, and the inferred population of low-mass stars are shown as small black points.
  }\label{fig:StarPositionsSubcluster}
\end{figure}

In figures \ref{fig:StarPositionsAll} and \ref{fig:StarPositionsSubcluster} we show the positions of stars in our model of DR21 in the xy plane (equivalent to the RA-Dec plane of the sky). 
Observed stars are shown as red circles and the inferred low-mass stars are shown as small black points. 

\begin{figure}
    \includegraphics[width=\columnwidth]{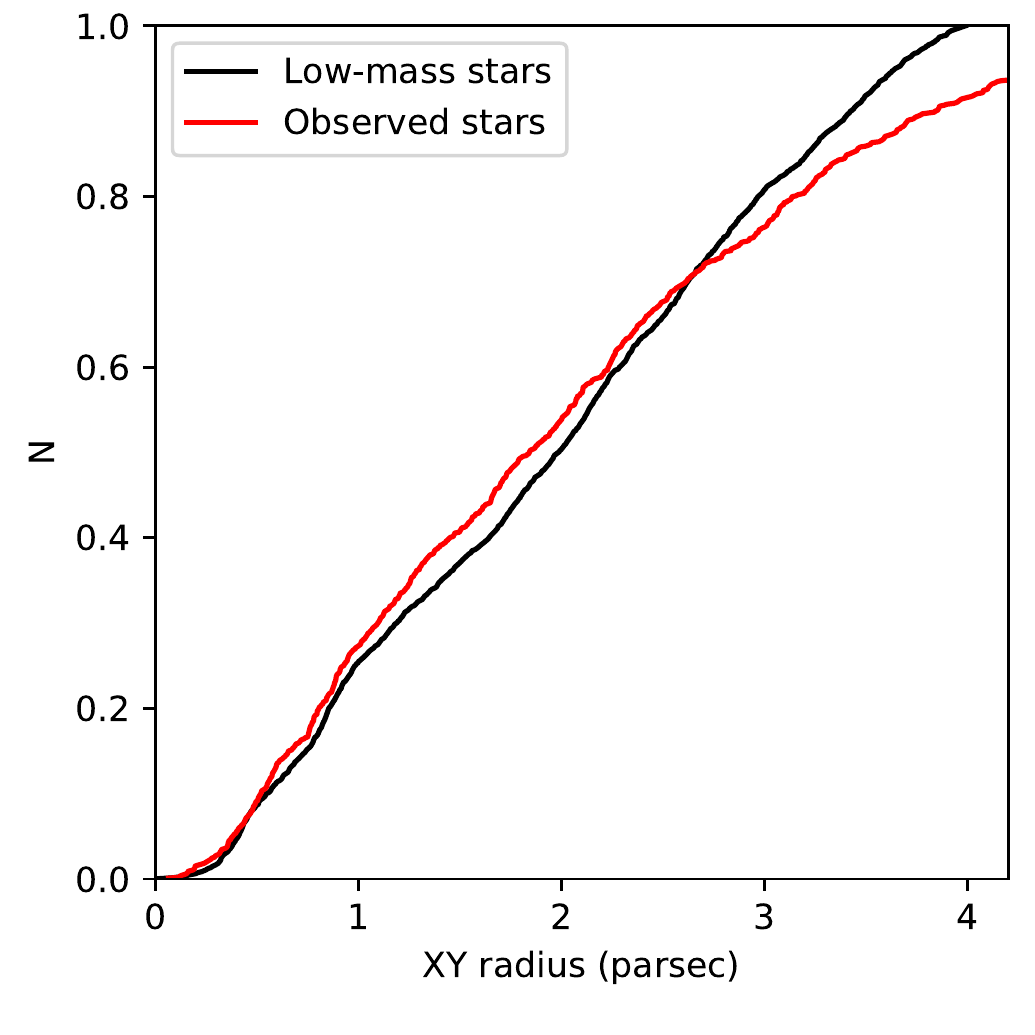}
    \caption{Cumulative radial distribution in the xy plane (i.e.\ the plane of the sky) for the initial positions of the stars in the DR21 simulation, measured from the centre of mass of our system. The red line shows the observed stars from the MYStIX sample while the black line shows the inferred low mass stars.
    }\label{fig:StarRadialDistribution}
\end{figure}

Figure \ref{fig:StarPositionsSubcluster} is a zoom-in of the most central sub-cluster, showing the subclustering of the low-mass stars and their distribution in an ellipse determined by the observed stars. 
To show the stellar distributions in a slightly different way, figure \ref{fig:StarRadialDistribution} shows the cumulative radial distributions of the initial positions of the observed high mass and inferred low mass stars in DR21, measured in the plane of the sky from the centre of mass of our system. 
The radial distributions of the two populations are very similar until the edge of our simulation area.

\begin{figure}
	\includegraphics[width=\columnwidth]{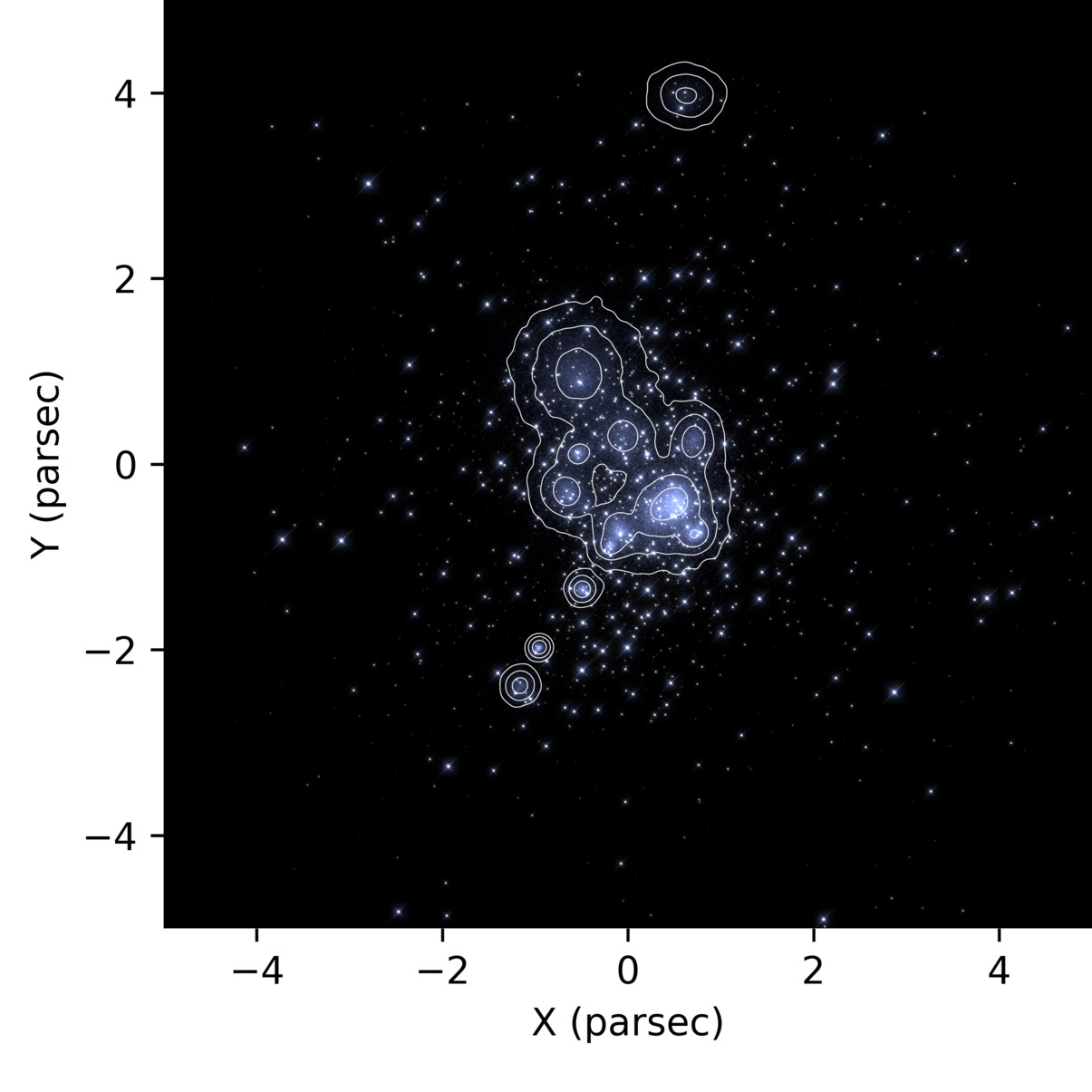}
    \caption{Initial conditions for M17.
    See figure~\ref{fig:DR21initial} for a description.
    The lowest contour level corresponds to a surface density of $167~\MSun~\pc^{-2}$, and each subsequent line is a factor of $10^{0.5}$ higher.
    }\label{fig:M17initial}
\end{figure}

\subsubsection{Velocities}\label{sec:initial-velocities}
We do not have observed velocity information for the stars in our sample.
Therefore, this is a parameter that we vary during our simulations.
In general, we give the stars in each sub-cluster a velocity dispersion which is related to the virial parameter $\alpha_{\rm vir}=2T/|W|$.
We note that this definition of the virial parameter typically is calculated using only the potential of the stars, and not of the gas which is also present in the region.
Therefore, it does not reflect the true virial state of the system.
However, most observational studies which estimate the velocity in star forming regions from virial arguments make the same kind of assumption -- papers interested in the stellar velocities usually only include the stellar mass; papers interested in the gas properties only consider the gas mass.
When both stars and gas are considered \citep{Foster2015}, the stars are found to be moving with sub-virial velocities while the gas is moving with velocities closer to that set by virial equilibrium of the entire system.
Typical observed values of the velocity dispersions are in the range of a few \kms~\citep{Schneider2010, Foster2015}.
We investigated three values for $\alpha_{\rm vir}$ in our simulations: virial systems with $\alpha_{\rm vir} = 1$, a subvirial system with $\alpha_{\rm vir} = 0.08$, and a supervirial system with $\alpha_{\rm vir} = 2$ (again, where we use only the kinetic and potential energy of the stars, and neglect the contribution of the gas to the gravitational potential). 

We give the unclustered stars an overall velocity dispersion as well, drawn from a user-specified value ($\sigma_{\rm stars}$ = 0-2 \kms).
In all cases, we give each star in its component (sub-cluster or unclustered background) a velocity vector with three random components, and then ensure that the dispersion of each grouping of stars has the required value. 

We can also allow each sub-cluster to have a bulk velocity relative to the others in the simulation, $V_{\rm rel}$.
We implement this by allowing each sub-cluster to have a velocity whose magnitude is set to a user-specified value, but the direction is randomized except that the y velocity must be acting along the gas filament if one exists, and the x and z velocities have smaller magnitudes that the y velocity. In these simulations, we investigated velocities up to 2 \kms, comparable to the velocity dispersions within the sub-clusters. 

\subsubsection{Masses}\label{sec:initial-masses}
We have measured masses for a small number of stars in the MYStIX sample (Kuhn, private communication), but the mass information is far from complete.
Therefore, we randomly assign masses from a broken power law IMF \citep{Kroupa2001} between 0.83 and 10 \MSun\ for the stars with observed positions, and randomly assign masses from the same IMF between 0.08 and 0.83 \MSun\ for the unobserved low mass stars.
For simplicity, we assume that all the stars in our simulation are single. 

\subsubsection{Gas}\label{sec:initial-gas}
After the stellar positions are determined, we add gas to the simulation.
Each sub-cluster is assigned its own ellipsoid of gas, with the same properties as the ellipsoids of stars -- it is a stretched Plummer sphere, with the same axis ratio and position angle as the stars.
We also include a sphere of gas to follow the unclustered stars.
The total mass in gas for each sub-region is set to be a fraction $f_{\rm gas}$ of the total mass in stars.
Our fiducial runs have the same gas and stellar masses, and we vary this ratio to determine the impact of gas mass on the resultant star cluster.
Finally we add a background of gas, a Plummer distribution with a given scale radius $r_a$ which can be either a sphere or a ``filament'', approximated by stretching the Plummer distribution into an ellipsoid with a very large axis ratio.
The mass of this background gas is loosely estimated from observations of the region of interest (\citet{Schneider2010} for DR21, \citet{Massi2015} for NGC~6357, and \citet{Reid2006} for M17), after accounting for the gas we have associated with the clustered and unclustered stars.
We note that the choices of gas mass for star-forming regions should be considered representative rather than precise, as the observations from which gas masses are determined usually cover a larger region than the MYStIX stellar observations, and the structure of the gas is complicated (holes, bubbles, filaments, etc.) rather than the smooth distribution we have assumed in our simulation.
All gas particles have the same mass (0.05 \MSun\ in the simulations presented here).

Our gas has an initial temperature of 10 K, consistent with temperature determinations of the typical gas temperature in star-forming regions such as DR21 \citep[e.g.][]{Schneider2010}.
The gas temperature is observed to reach 60-100 K in the centres of clumps, likely due to heating from the natal protostars.
In our code, however, we treat our gas with an adiabatic equation of state and do not include radiative heating or cooling, so a uniform temperature for the gas is a reasonable but simplistic approximation.
Currently these simulations do not include sink particles, so we cannot follow star formation from the collapse of gas to very high densities.
We give the gas a velocity dispersion, $\sigma_{\rm gas}$, calculated within each sub-cluster or the background, which has the same value for all regions.  

\subsubsection{Regions}\label{sec:initial-regions}
For our parameter space study, we concentrated on the DR21 region.
This is one of the youngest regions in the MYStIX sample, and is deeply embedded.
It lies in one end of a long, dense molecular filament, and there are a number of sub-filaments outside the MYStIX region which are approximately perpendicular to the main filament.
We chose this region for our main investigation since it appears to be in an earlier stage of formation than most of the MYStIX regions.
We also ran a simulation for M17 and NGC~6357 with our default parameters, to look at the effect of different morphologies on our results.
M17 is an older region which is mostly spherical, but still contains some substructure within the main cluster.
In the MYStIX classification, it is considered to have a `core-halo' structure.
There is still a fair amount of gas within the M17 star cluster region, suggesting it still has some time before it completely emerges from the natal cloud.
NGC~6357 consists of three main regions, each with a core-halo structure.
They are located on or near the edges of a large bubble that has been blown in the ISM some time ago.

\begin{figure}
	\includegraphics[width=\columnwidth]{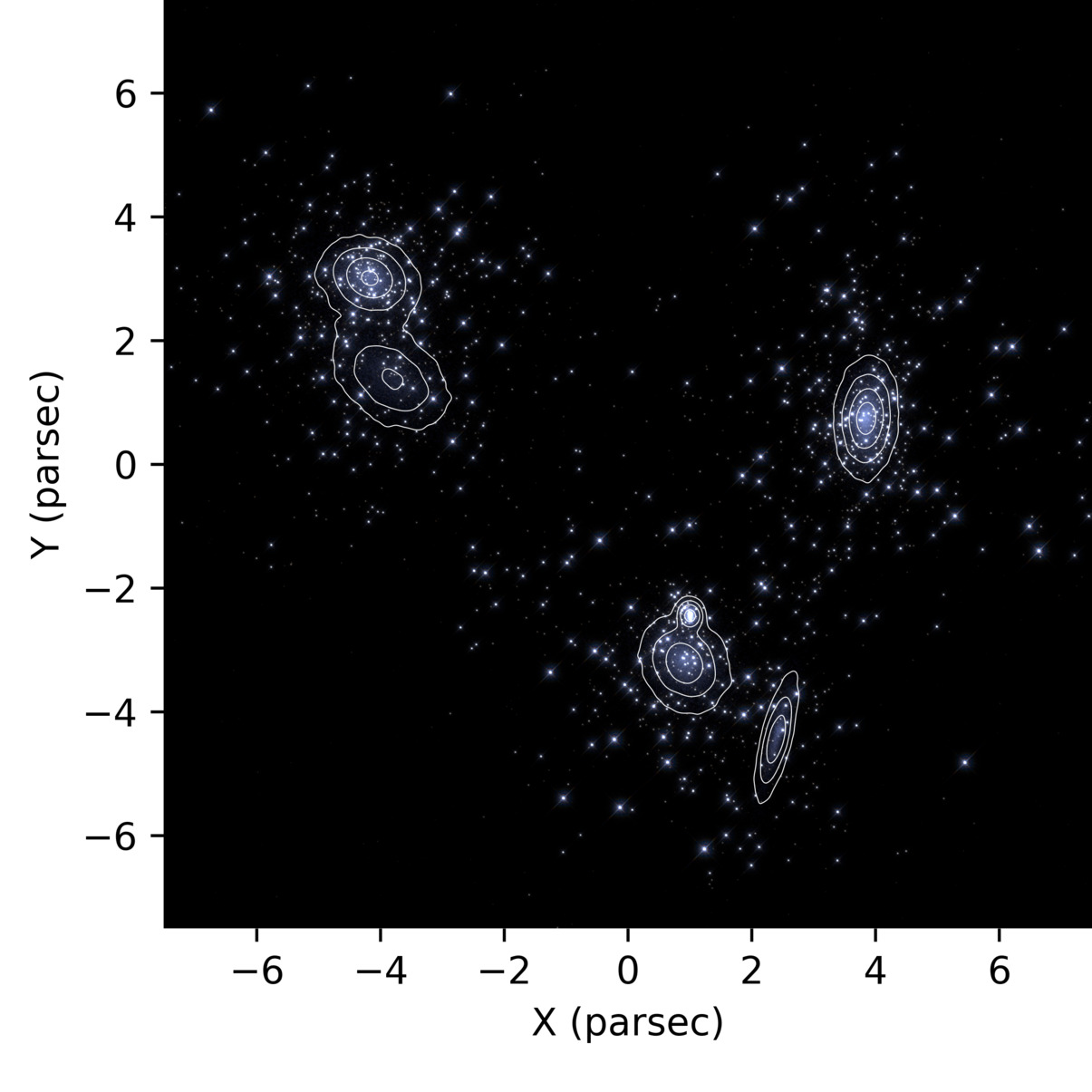}
    \caption{Initial conditions for NGC~6357.
    See figure~\ref{fig:DR21initial} for a description.
    The lowest contour level corresponds to a surface density of $103~\MSun~\pc^{-2}$, and each subsequent line is a factor of $10^{0.5}$ higher.
    }\label{fig:NGC6357initial}
\end{figure}

The initial conditions for the DR21 simulation, as well as for two other regions that we modelled (M17 and NGC~6357) are shown in figures~\ref{fig:DR21initial},~\ref{fig:M17initial}, and~\ref{fig:NGC6357initial}.
These images show the xy plane (i.e.\ the plane of the sky).
The stars are represented as objects with intensity and colour related to their mass and radius.
Gas is rendered as diffuse emission and its smoothed surface density is also shown with contour lines.

\begin{table*}
	\centering
	\caption{Cluster Parameters}\label{tab:clusters}
    \begin{tabular}{lcccccccc}
		\hline
    	Cluster & RA & Dec & Distance & \multicolumn{3}{c} {Background Gas} & Extent of unclustered & Total number\\
        & (deg) & (deg) & (\pc) & Mass & r$_a$  & ellipticity & stars $z_{\max}$ & of stars \\ 
    	\hline
    	DR21     & 309.752083 &  42.345092 & 1500 & 1500 \MSun & 10 \pc & 0.8 &  4 \pc & 3994 \\
    	M17      & 275.099269 & -16.174685 & 2000 &  800 \MSun & 10 \pc & 0   &  5 \pc & 17870 \\
    	NGC~6357 & 261.352832 & -34.331317 & 1700 & 2000 \MSun & 25 \pc & 0   & 12 \pc & 15078 \\
    	\hline
    \end{tabular}
\end{table*}
\begin{table*}
	\centering
	\caption{Simulation Parameters: Stellar virial parameter ($\alpha_{\rm vir}$), velocity dispersions of unclustered stars and gas ($\sigma_{\rm stars}$ and $\sigma_{\rm gas}$), fraction of gas in each sub-cluster relative to stellar mass ($f_{\rm gas}$), maximum velocity of the subclusters relative to each other ($V_{\rm rel}$), and maximum offset of each subcluster from the z=0 plane ($Z_{\rm offset}$).
    }\label{tab:parameters}
	\begin{tabular}{lcccccc} 
		\hline
		Model Name & $\alpha_{\rm vir}$ & $\sigma_{\rm stars}$ & $\sigma_{\rm gas}$ & $f_{\rm gas}$ & $V_{\rm rel}$ & $Z_{\rm offset}$\\
		\hline
		fiducial & 1 & 1 \kms & 1 \kms & 1 & 0 \kms & 0 \\
		subvirial & 0.08 &  1 \kms & 1 \kms & 1 & 0 \kms & 0 \\
		supervirial & 2 &  1 \kms & 1 \kms & 1 & 0 \kms & 0 \\
		low stellar sigma &  1 & 0 \kms & 1 \kms & 1 & 0 \kms & 0 \\
        high stellar sigma & 1 & 2 \kms & 1 \kms & 1 & 0 \kms & 0 \\
        low gas sigma & 1 & 1 \kms & 0 \kms & 1 & 0 \kms & 0 \\
        high gas sigma & 1 & 1 \kms & 2 \kms & 1 & 0 \kms & 0 \\
        low gas fraction & 1 & 1 \kms & 1 \kms & 0.1 & 0 \kms & 0 \\
        high gas fraction & 1 & 1 \kms & 1 \kms & 10 & 0 \kms & 0 \\
        middle relative velocity & 1 & 1 \kms & 1 \kms & 1 & 1 \kms & 0 \\
        high relative velocity & 1 & 1 \kms & 1 \kms & 1 & 2 \kms & 0 \\
        middle offset & 1 & 1 \kms & 1 \kms & 1 & 0 \kms & $0.5\times z_{\max}$\\
        high offset & 1 & 1 \kms & 1 \kms & 1 & 0 \kms & $1.0\times z_{\max}$\\
        \hline
	\end{tabular}
\end{table*}

In table~\ref{tab:clusters} we give the global information about each of our modelled regions (DR21, M17 and NGC~6357).
These do not change when we were doing our parameter space study, and are based on the observed properties of the regions.
In table~\ref{tab:parameters} we give the simulation parameters that we varied for each run.
These include the virial parameter of each stellar sub-cluster ($Q$, defined using the kinetic and potential energy of the stars alone); the velocity dispersions $\sigma$ for the stars and for the gas; the gas fraction $f_{\rm gas}$, given as the ratio between the mass of gas to the mass of stars in each sub-cluster; the maximum relative velocity of the sub-clusters $V_{\rm rel}$; and the maximum distance of each sub-cluster from the z=0 plane $z_{\max}$.
We performed the fiducial simulations for all three regions; the other simulations were done using DR21 as the test region.
       
\subsection{Evolution}

After we have created our initial conditions, we use AMUSE to evolve the system forward in time.
For these particular simulations, we use the 4th-order Hermite gravitational N-body code {\tt ph4}, purpose-written for AMUSE, to evolve the stars under the influence of their own self-gravity.
We use the smoothed particle hydrodynamics code {\tt Gadget2} \citep{Springel2005} to evolve the gas under their own self-gravity and hydrodynamic forces.
To couple the gas to the stars, we use the Bridge formalism \citep{Fujii2007} with the dynamics code {\tt FastKick}.
This mechanism allows us to self-consistently model the motion of all components of the system according to the full gravitational potential, while still using the most appropriate numerical techniques for the individual sub-systems given the natural times on which they evolve.
Bridge works by having each system experience the gravity of the other through periodic velocity kicks, with normal evolution in between.
Particular attention must be paid to the relative timesteps of these drift-kick interactions, and we use a short interaction time of 0.01 N-body system times in all our simulations, as our gas and stars are physically intermixed.

For simplicity, we assume that our gas is appropriately modelled with an adiabatic ideal gas equation of state.
We do not include star formation or stellar feedback, either in the form of stellar winds or of supernovae.
Therefore, we limit our simulations to the first 10 Myr of the clusters' evolution, as that is below the lifetime of the most massive stars in our simulation. 

\section{Results}\label{sec:result}
\begin{figure}
  \centering
	\includegraphics[width=0.31\columnwidth]{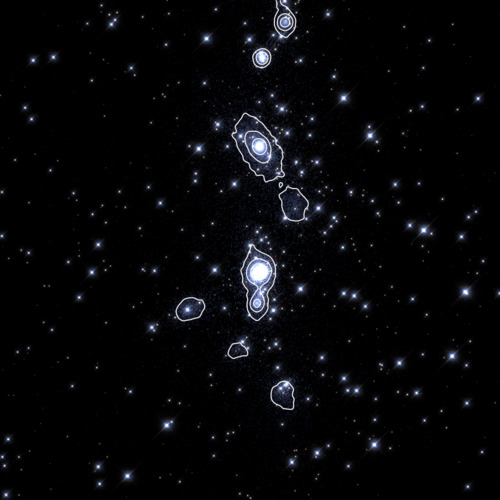}
	\includegraphics[width=0.31\columnwidth]{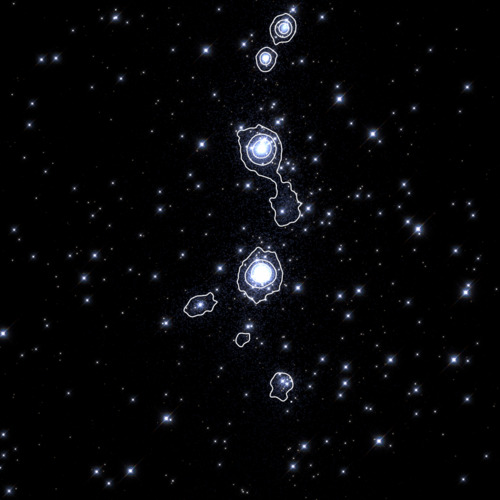}
	\includegraphics[width=0.31\columnwidth]{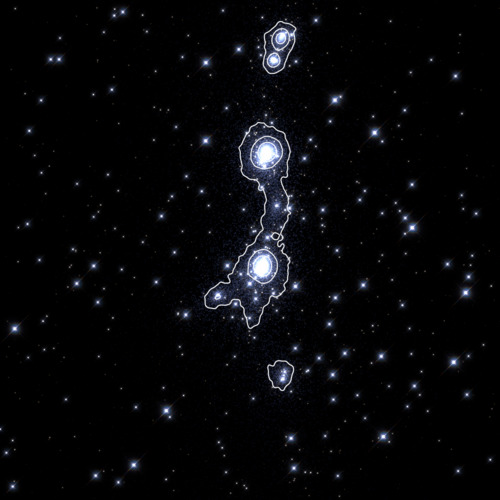}\\
	\includegraphics[width=0.31\columnwidth]{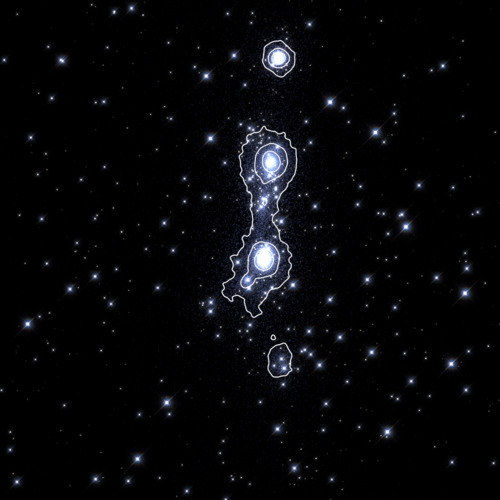}
	\includegraphics[width=0.31\columnwidth]{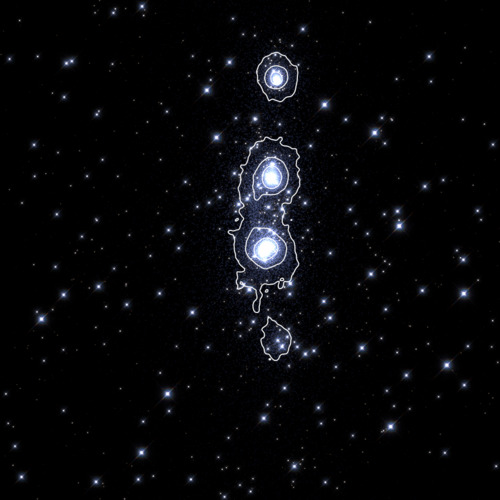}
	\includegraphics[width=0.31\columnwidth]{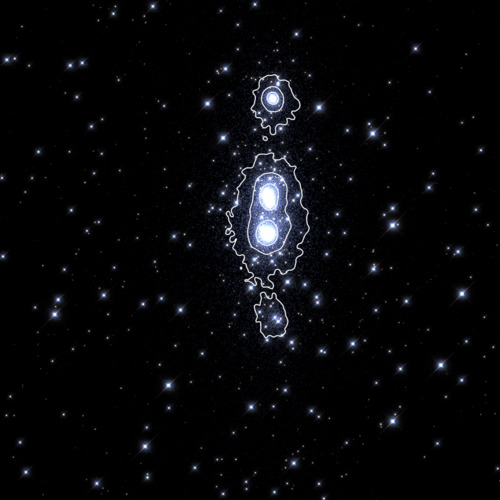}\\
	\includegraphics[width=0.31\columnwidth]{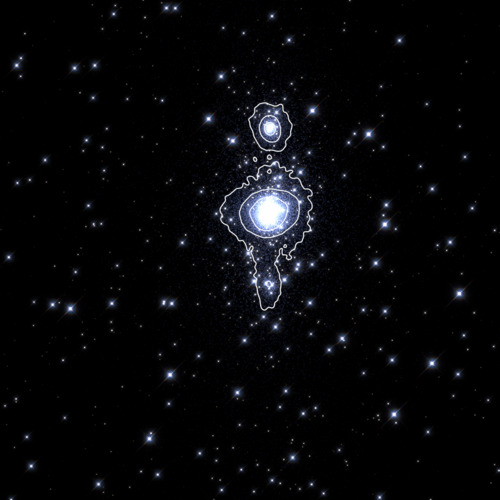}
	\includegraphics[width=0.31\columnwidth]{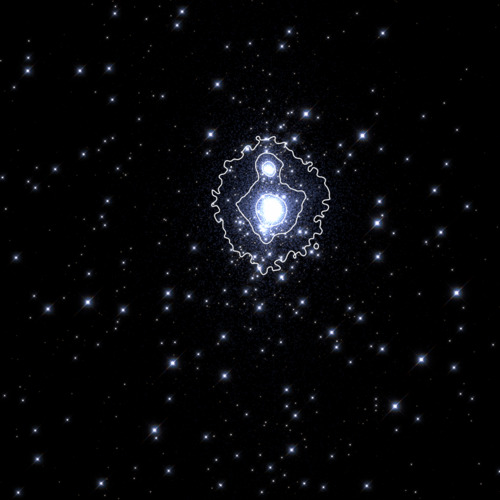}
	\includegraphics[width=0.31\columnwidth]{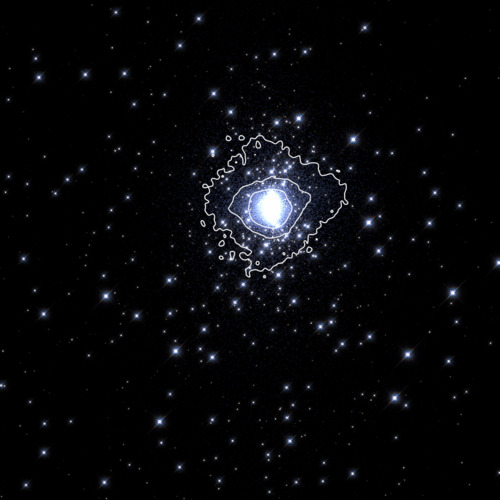}\\
  \caption{Snapshots of the first 9 snapshots of the evolution of DR21 in our fiducial simulation, from top left to lower right.
    The snapshots are taken 100 000 years apart, with the first snapshot 100 000 years after the simulation begins.
    Each snapshot is 5 \pc~on a side, otherwise as in figure~\ref{fig:DR21initial}.
    The subclusters have almost entirely merged within the first million years of cluster evolution, and the gas density has become quite spherical.}
    \label{fig:DR21evolution}
\end{figure}

All our simulated systems quickly collapse to a nearly spherical, centrally concentrated cluster, as shown in figure~\ref{fig:DR21evolution} for the fiducial DR21 simulation.
The timescale for this collapse is short, on the order of a Myr, and the timescale does not depend much on the choices of initial parameters.
This result is in agreement with N-body, gas-free simulations of subclustered regions \citep[e.g.][]{Fellhauer2009, Fujii2012, Parker2016}, where the clusters lose their substructure on approximately the same timescale.
It is interesting to note that in the \citet{Parker2016} simulations, the systems with subvirial velocity dispersions evolve to smooth clusters more quickly than those with a higher initial virial parameter $\alpha_{\rm vir}=2T/|W|$.
Our simulations have an additional gravitational potential from the gas mass, which effectively decreases the virial parameter and shortens the time for collapse. 

\subsection{Substructure}\label{sec:result-structure}

To characterize the substructure seen in our simulations, we used the $\mathcal{Q}^{+}$ algorithm \citep{Jaffa2017}, which is not related to the virial parameter but rather characterizes the fractal distribution of the stars.
The $\mathcal{Q}^{+}$ algorithm describes a sub-structured star cluster in terms of three parameters: the fractal dimension, the number of levels, and the density scaling exponent.
The fractal dimension describes how clumpy or smooth the distribution is, with 1 being very clumpy and 3 being smooth.
The number of levels describes the range of scales over which the fractal sub-structure exists, with a low value indicating that the smallest sub-structures are not much smaller than the overall cloud and a high value indicating that the smallest structures are very small compared to the cluster size.
The density scaling exponent describes where most of the mass is distributed in the hierarchy, with a high value indicating that most of the mass is confined to the smallest sub-structures and a low value indicating that the sub-clumps are not much more dense than the larger clumps. Clusters that are not sub-structured but centrally condensed can be characterised by a radial density exponent using the method described in \cite{Cartwright+2004} which is incorporated into the $\mathcal{Q}^{+}$ analysis code.

What we found was that the fractal dimension of the DR21 fiducial simulation was consistently 1.58 (quite clumpy) until the sub-clusters merged at around 0.5 Myr.
However, during that time, the number of levels decreases steadily and the density scaling exponent increases, suggesting that the stars from the smallest sub-clump(s) are spreading out to fill the larger hierarchy.
After the clumps merge at $\sim 0.5$ Myr, the region completely loses its fractal structure and is consistent with a centrally concentrated system.
The central stellar density increases with time, as does the radial density gradient. 

\begin{figure}
    \includegraphics[width=\columnwidth]{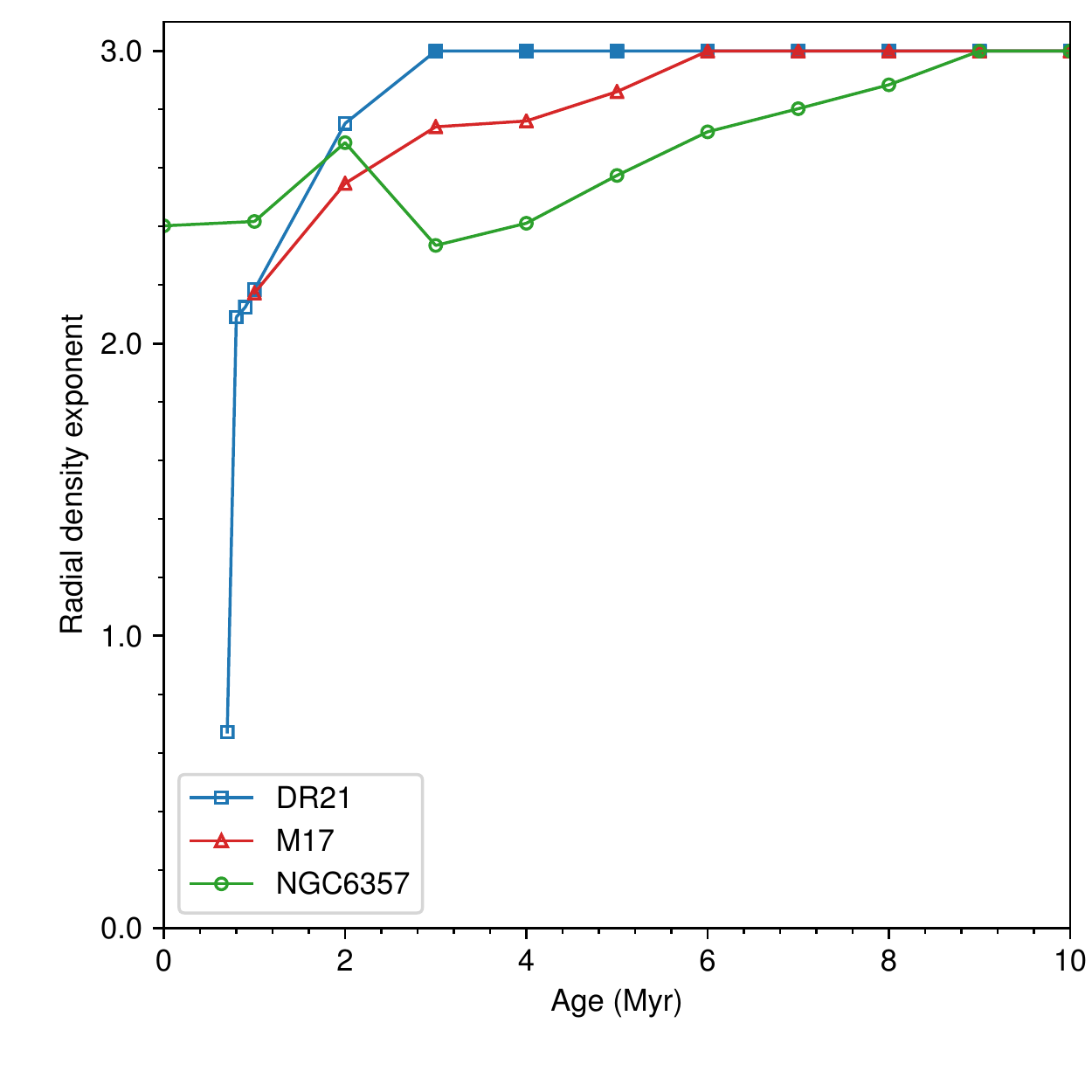}
	\caption{Figure showing the evolution of the cluster structure over time.
    We analyse only the massive stars taken from the MYStIX observations, as this should be similar to what observers would be able to detect.
    Filled symbols indicate lower limits ($\alpha \geq 3$) as the analysis we perform cannot estimate radial density exponents greater than 3.
    The lines begin at the timestep in which the fractal structure is lost and the cluster becomes classified as centrally concentrated.
    }
    \label{fig:fractal_parameter_evolution}
\end{figure}

If we limit our analysis to only the high mass stars (${\rm M} > 0.83~\MSun$) that were taken from the MYStIX observations and exclude the low mass stars added for completeness we can compare the structure that would be theoretically seen by an observer to the full structure we have in the simulations.
These results are shown in figure~\ref{fig:fractal_parameter_evolution}.
For the initial conditions of DR21 this mass limited analysis gave a lower fractal dimension and density scaling exponent than analysis of the full cluster but a similar number of levels.
This suggests that the observable higher mass stars populate a more clumpy distribution, but are not confined to the smallest sub-clumps.
Their evolution showed the same trend as the full cluster.
The high mass stars in M17 and NGC~6357 exhibit similar behaviour; the initial conditions have a low fractal dimension and high number of levels, but the substructure is quickly erased (within 1-2 Myr) and the cluster becomes centrally concentrated.
It is interesting to note that when the full clusters are analysed, M17 would not be classified as fractally sub-structured and NGC~6357 only just so.
It is only when we analyse the high mass component of these two clusters that the sub-structure is detected. However, this difference in structure is ambiguous and may be caused by limitations in the $\mathcal{Q}^{+}$ analysis rather than an actual difference in structure between the high mass and low mass components.

\subsection{Structural Properties}\label{sec:results-structure}

The MYStIX observations provide a large sample of structural properties of embedded stellar (sub-)clusters, to which we can compare our models.
In order to make this comparison, we measure structural properties of our simulations every $\sim$100 000 years for the first 7~Myr of their evolution.
We treated each snapshot as an `independent observation' in order to make the comparison to the MYStIX observations.
The MYStIX regions have a variety of ages, masses, and (most probably) initial conditions, so a one-to-one comparison is not straight-forward.
However, by looking at the ensemble properties of the model snapshots compared to the observations, we are able to make inferences about the evolution of these regions and the influence of the various values of the initial parameters we chose for our simulations. 

\begin{figure}
	\includegraphics[width=\columnwidth]{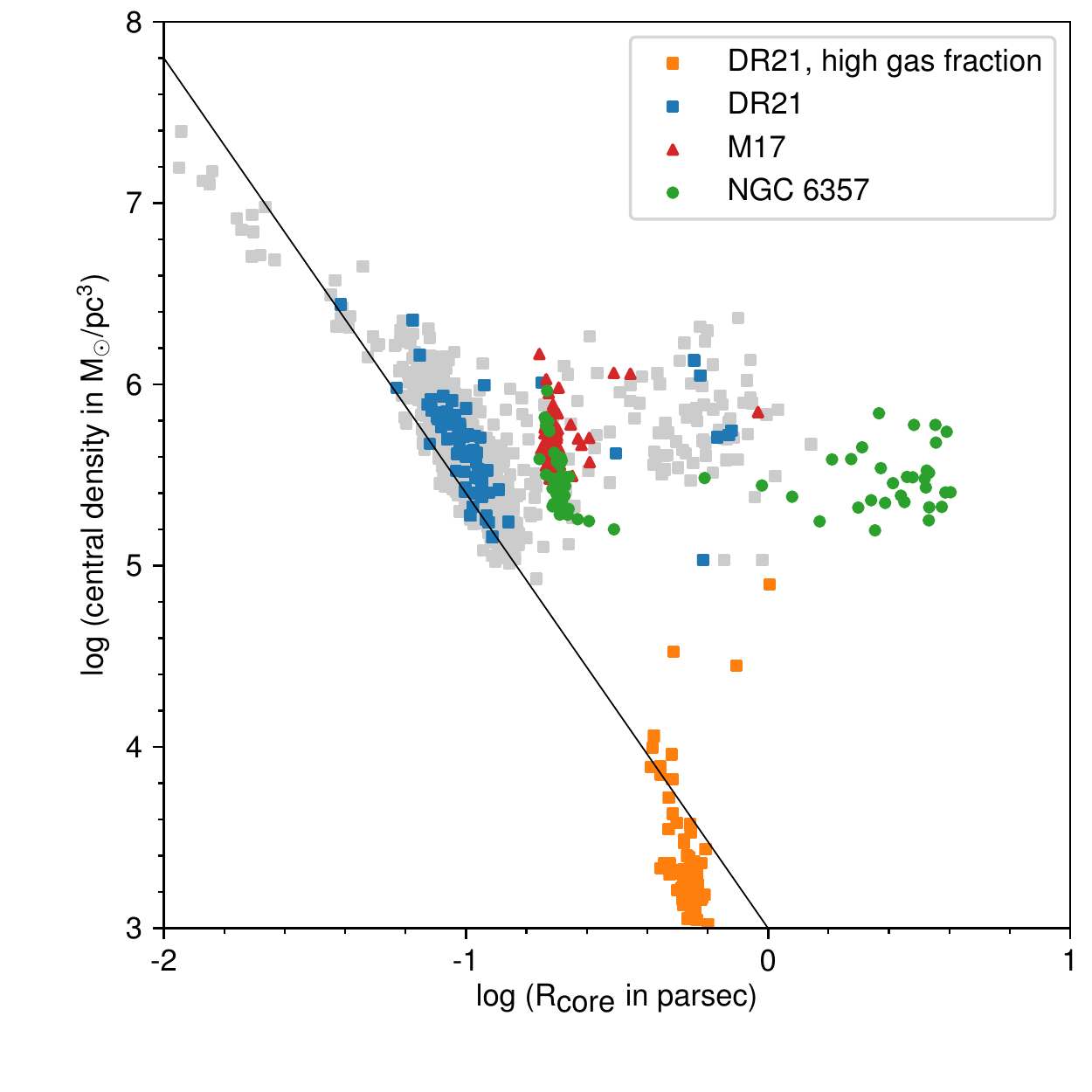}
    \caption{Central stellar density vs core radius, for all runs.
      The first 7 Myr of evolution was saved in snapshots every $\sim$ 100 000 yr, then each density-radius pair plotted as if it was a unique observation.
      The solid line is best-fit to MYStIX observations.
      The various symbols and colours correspond to different simulations.
      Four of importance are the DR21 fiducial simulation (blue squares), the DR21 simulation with high gas fraction (orange squares), the M17 fiducial simulation (red triangles), and the NGC~6357 fiducial simulation (green circles).
      The grey squares are variations of the DR21 simulation with the parameters given in table~\ref{tab:parameters}.
    }\label{fig:density_radius}
\end{figure}

The two main structural parameters that are best measured in the MYStIX sample are the core radius of the subclusters and their central density of stars.
The MYStIX data show a clear anti-correlation between the two quantities, shown as a solid line in figure~\ref{fig:density_radius}.
Our models all begin with core radii that are much larger than the observed sub-clusters (to the right part of figure~\ref{fig:density_radius}), which is simply a result of our analysis -- we calculate a single core radius for each snapshot, and so when the simulations still maintain their sub-structured appearance, this core radius will be too large.
Our models evolve first to smaller core radii at approximately constant density, and then evolve along the MYStIX relation.
This agreement suggests that the central density-core radius relationship is fundamental to the structure of young stellar clusters. 

We also point out that the various parameters that we chose to investigate in our models make, for the most part, very little difference to the overall structure and evolution of the system.
To guide the reader's eye, we highlight four particular simulations in figure~\ref{fig:density_radius}.
First, our canonical model, the DR21 fiducial simulation is shown as blue filled squares.
The initial conditions have an overall core radius around 1~\pc~and a central density of $\sim 10^6~\MSun/\pc^3$, and after about 0.5~Myr, the model has come close to finishing its collapse to a single, spherical system with a core radius of about 0.1~\pc.

Almost all the other DR21 models show a very similar evolution, with some variation in the exact values of core radius and central density.
The clear exception is shown in orange squares.
This is the simulation that begins with significantly more gas -- a factor of 10 more mass in gas than in stars in each of the initial subclusters.
Only in this case does the stellar density remain significantly lower (below $\sim 10^6~\MSun/\pc^3$), but still the simulations evolve quickly to the observed relation.
Interestingly, the opposite simulation with a factor of 10 less gas mass than stellar mass does not show as significant a difference from the fiducial simulation, suggesting there is a threshold required before the gas mass has an observable effect on the stellar structure. 

\begin{figure}
	\includegraphics[width=\columnwidth]{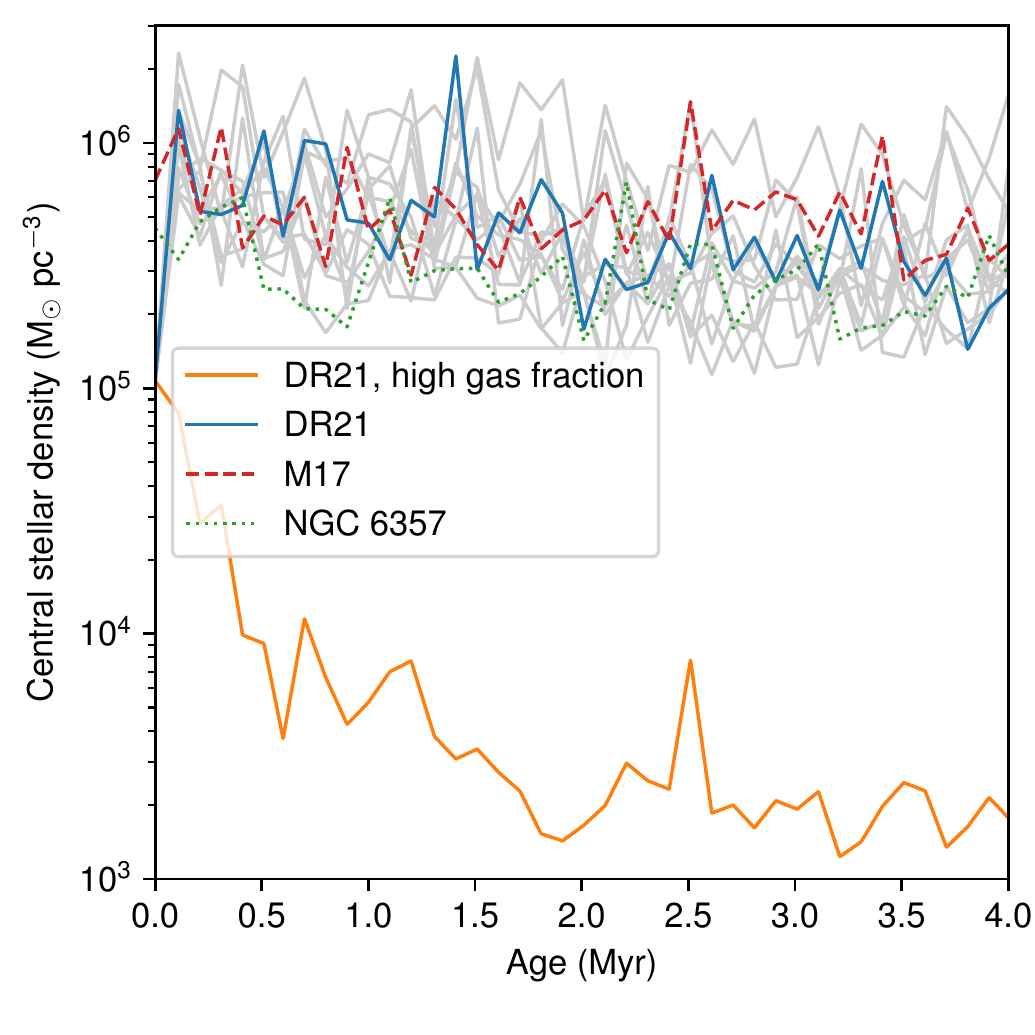}
    \caption{Central stellar density as a function of time, for all runs.
    The different colours correspond to the same simulations as in figure~\ref{fig:density_radius}.
    }\label{fig:stellardensity}
\end{figure}

The other two simulations to note are the two fiducial simulations for the other two regions (M17 -- red triangles, and NGC~6357 -- green circles).
Despite the difference in the total mass in stars, and their initial sizes, these simulations also follow approximately the same evolution in this diagram as the DR21 simulations. 
Because NGC~6357 is initially much larger (core radius $\sim$ 3~\pc~or so), it takes longer to collapse, but otherwise the evolution and its endpoint is the same. We note that the MYStIX observations show a scatter of about an order of magnitude in central density at a given core radius (above the line drawn in figure \ref{fig:density_radius}) and that the observed subclusters in DR21 lie along the bottom of that scatter while the subclusters in M17 and NGC 6357 lie closer to the top of the scatter. 
This fundamental relationship between core radius and central density is a natural outcome of dynamical evolution of merging stellar sub-clusters. To facilitate the interpretation of the previous figure, we include the evolution of the stellar density as a function of time in figure \ref{fig:stellardensity}. The line colours are the same as the colours shown in figure \ref{fig:density_radius}. 

\begin{figure}
	\includegraphics[width=\columnwidth]{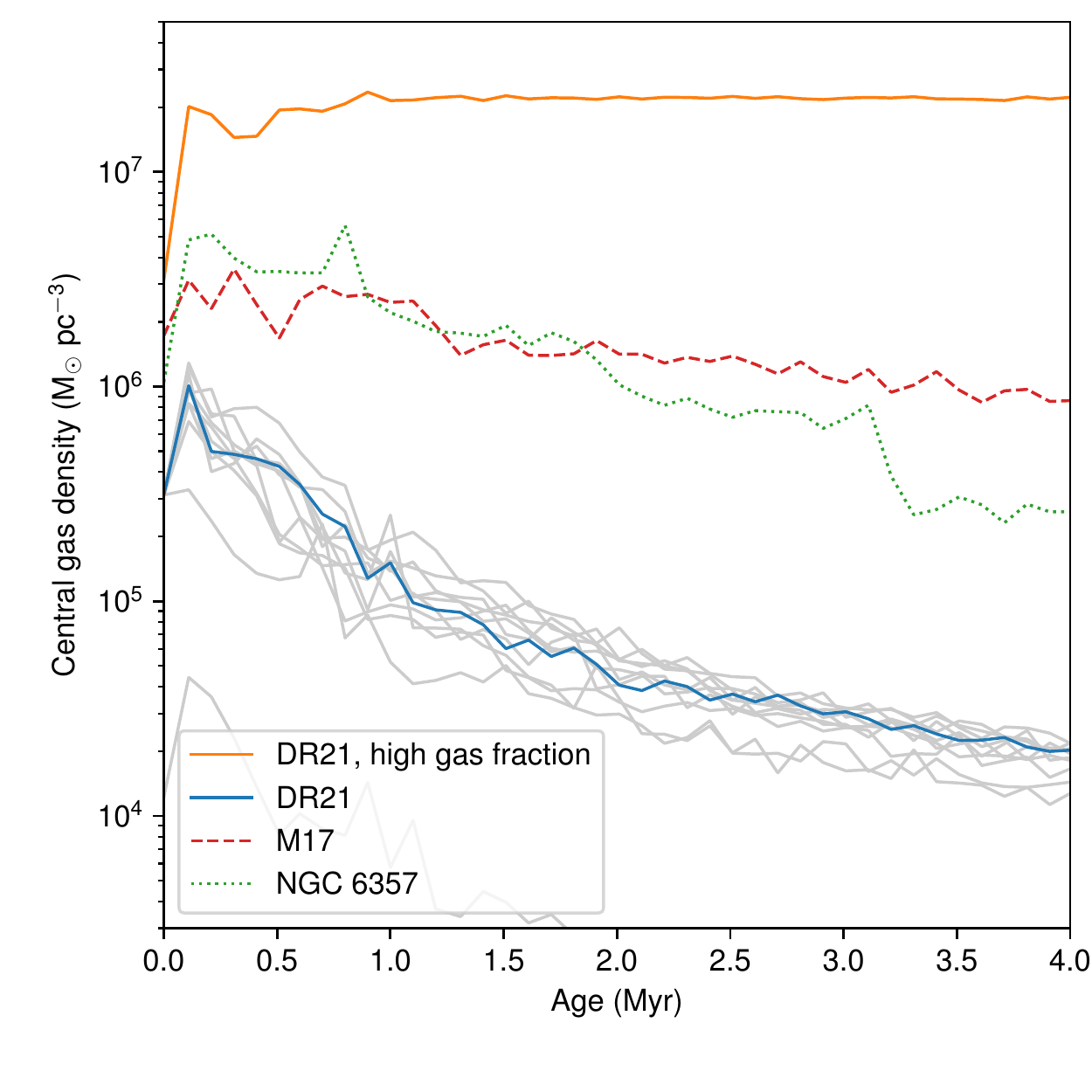}
    \caption{Central gas density as a function of time, for all runs.
    The different colours correspond to the same simulations as in figure~\ref{fig:density_radius}.
  }\label{fig:gasdensity}
\end{figure}

In contrast to the stellar density of the cluster, the central gas density decreases with time, as shown in figure~\ref{fig:gasdensity}. 
By looking at the half-mass radius or virial radius of the gas, we can see that as the stellar cluster contracts and the sub-clusters merge, the gas moves outwards.
At these early, gentle stages of gas motion, the stars shows no signs of reacting strongly to the reduction in the overall cluster potential -- rather than expanding, the stars continue to contract slowly over time. This is in contrast to simulations in which the gas potential is removed quickly or instantaneously, suggesting that the timescale of the gas removal mechanism is important for the subsequent evolution of the star cluster. 

\subsection{Kinematic Properties}
\label{sec:results-kinematic}

\begin{figure}
	\includegraphics[width=\columnwidth]{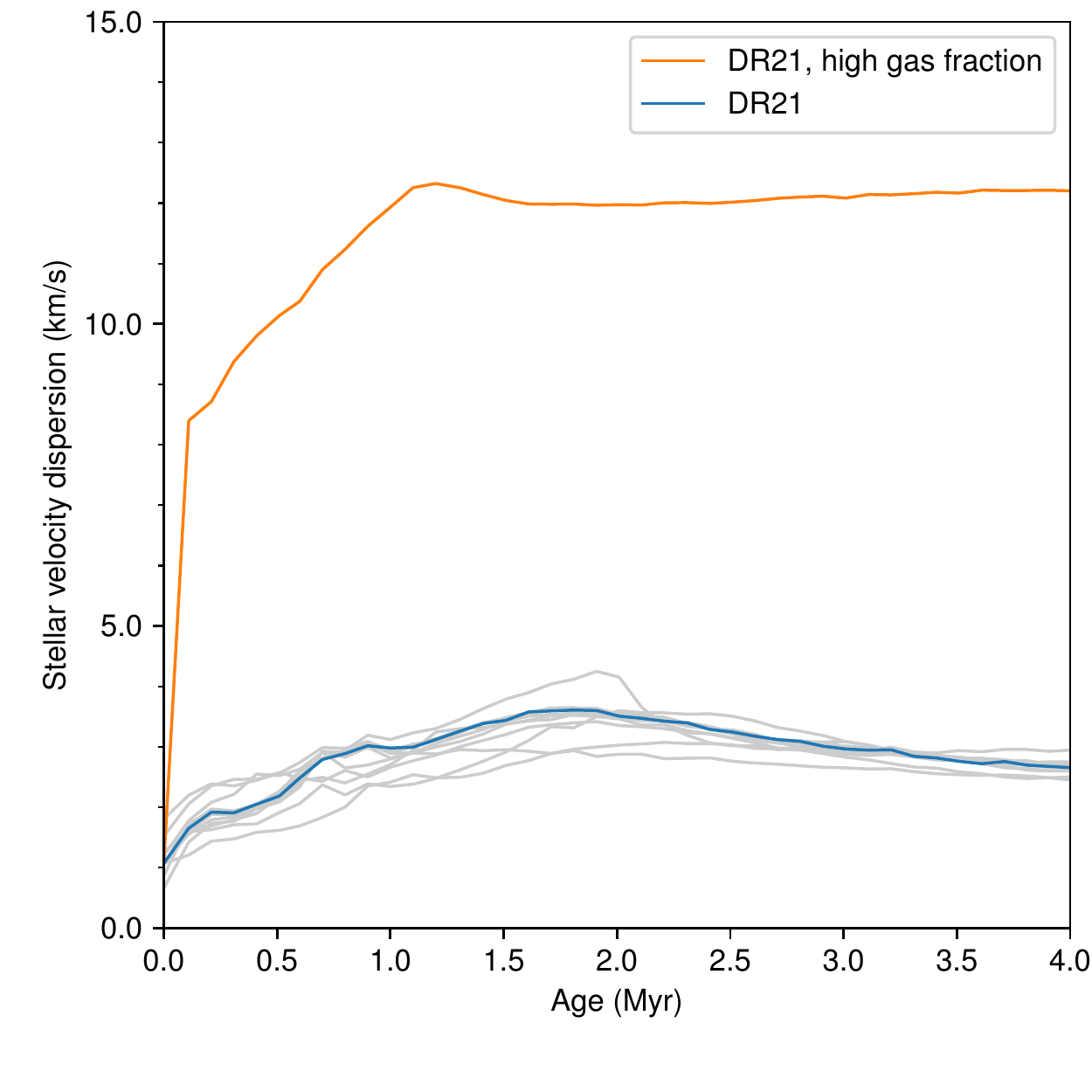}
    \caption{Stellar velocity dispersion as a function of time for all the DR21 region runs.
    The different colours correspond to the same simulations as in figure~\ref{fig:density_radius}.
    The orange line, with the higher velocity dispersion, is the high gas fraction simulation. 
    }\label{fig:sigma_time}
\end{figure}

Our expectation from previous simulations is that as the clusters collapse, the stars' velocity dispersion will increase \citep{Foster2015, Parker2016}.
Our simulations show the same behaviour, but only during the collapse phase.
Shortly afterwards, the clusters settle to a steady state appropriate for their new spherical configuration, as shown in figure~\ref{fig:sigma_time}.
Again, the various parameters we chose to investigate result in very little difference in the value or evolution of the velocity dispersion, except when the region has substantially more gas (orange line) than our fiducial run (blue line).
The velocity dispersion does not change substantially even as the region collapses.  Because we have all the information for each particle in our simulation, we can directly calculate the total kinetic and potential energies in the simulation, and therefore the virial ratio. Observationally, however, the virial ratio is often inferred from the velocity dispersion, estimates of total mass and radius of the region, and assumptions about density distributions.

\begin{figure}
    \includegraphics[width=\columnwidth]{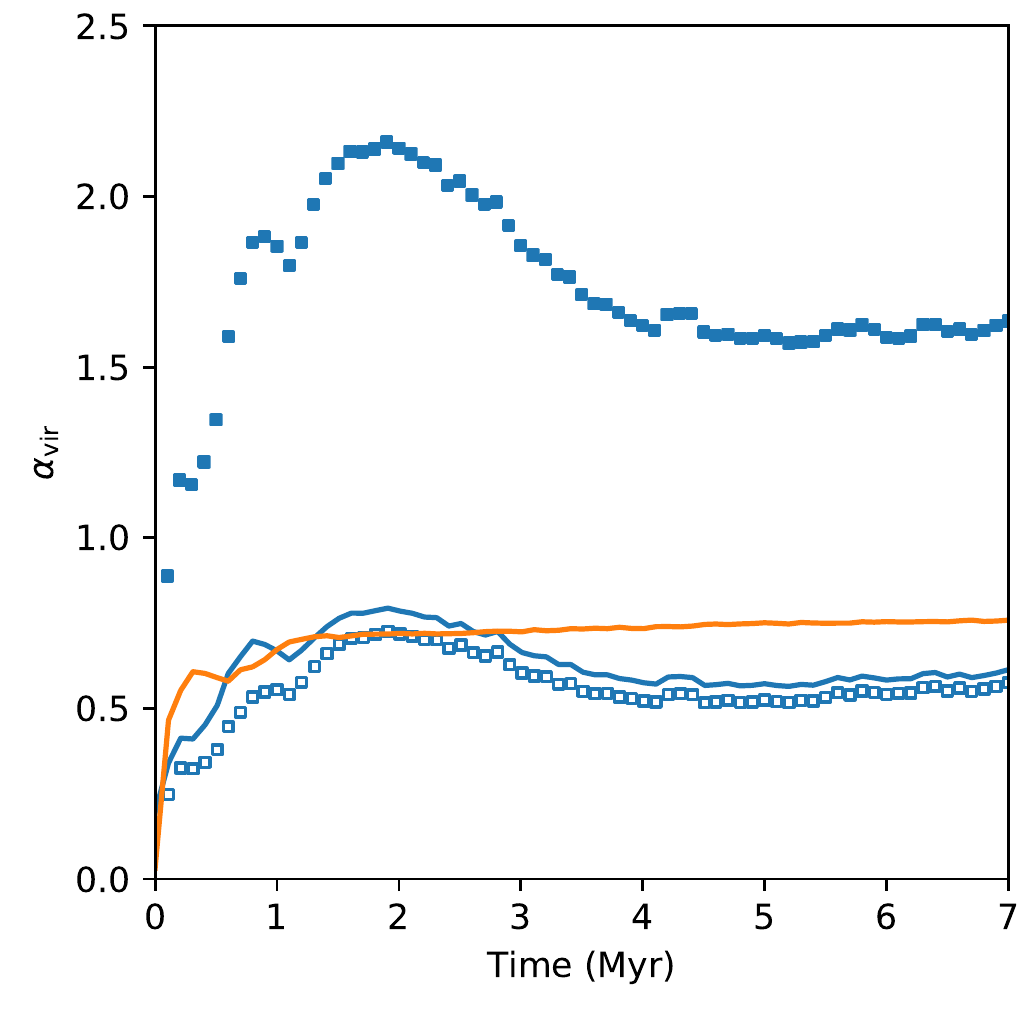}
    \caption{Virial ratio as a function of time for two DR21 region runs.
    The different colours correspond to the same simulations as in figure~\ref{fig:density_radius}.
    The orange line is the high gas fraction simulation. 
    The solid squares show the virial ratio in the fiducial DR21 simulation calculated using only the kinetic and potential energies of the stars; the open squares show the virial ratio for that simulation calculated using the kinetic energy of the stars but the potential energy of the gas and stars together.
    }\label{fig:virialratio}
\end{figure}

In figure~\ref{fig:virialratio} we plot the true virial ratio for the two simulations highlighted in the previous figure.
The high stellar velocity seen in the high gas mass simulation is caused by the increase in total gravitational potential, as both simulations have similar overall virial ratios.
The solid and open squares show other calculations of the virial ratio for the DR21 fiducial run. 
The solid squares use the information about stars alone without considering either the kinetic or the potential energy of gas.
Our gas mass and stellar mass are approximately the same in this simulation, and the gas has a very small kinetic energy.
Therefore, the stellar-only virial ratio is too large by a factor of about two, which could would us to an incorrect interpretation about the lack of boundedness of this region.
If careful observers were to consider the stellar kinetic energy and the potential of both the stars and the gas, they would calculate that the virial ratio was given by the open squares.
In this particular region, that value is much closer to the true value.
However, simply including all the mass contributions to the potential in a region may not be sufficient to corrected determine the virial ratio.
Embedded clusters and star-forming regions will have a different ratio of stellar to gas mass, and also stellar to gas velocity dispersion, throughout their evolution.
Observers must consider all contributions in order to understand the true virial state of these systems.

\section{Conclusions}

We have presented simulations of young, embedded star clusters in which we simultaneously model the dynamical evolution of both the stars and the gas, beginning with initial conditions that are motivated and directly drawn from observations.
By comparing our results to observations from the MYStIX regions, we conclude that our models match the observations well, and so the initial conditions and choices of {\it a priori}\ unknown parameters we used are realistic and reasonable.
We also conclude that the subsequent evolution of these regions is quite insensitive to reasonable values of the parameters that we varied.
We found that our initially very substructured systems became spherical, monolithic, and smooth very quickly, in both their spatial and velocity distributions.
Evolution of embedded clusters at young ages appears to be dominated by gravitational interactions between the stars, with only a small influence from the surrounding gas. 

Like all models, our simulations have limitations.
First, we neglect ongoing star formation, despite the high likelihood that it should happen in these regions.
At galactic scales, a typical star formation threshold used in simulations is $100~\MSun/\pc^3$ \citep[e.g.][]{Benincasa2016}, which our simulations regularly exceed at their centre.
However, simulations of cluster-scale regions suggest that a more reasonable star formation threshold is a few thousand $\MSun/\pc^3$ \citep{Myers2010} in star clusters, which is well below the maximum seen in almost all our simulations.
Only the simulation with the high gas fraction reaches about $10^5~\MSun/\pc^3$.
Star formation is also expected to proceed on timescales related to the local free-fall time \citep[e.g. $\sim$ 10 $\tau_{\rm ff}$,][]{Kruijssen2011}.
The free-fall time of the gas in the DR21 simulations is of order 1 Myr.
Since our total simulation time is at most 10 Myr, we feel justified in neglecting most star formation, especially over the early stages of the simulation.
However, to follow the evolution of young star clusters over a longer time or with a high gas fraction, active star formation should be included in the model. 

We also note that some of the MYStIX regions are embedded in gas filaments (particularly the DR21 region) in which the gas motions are directed along the filament towards the cluster.
Therefore, the gas mass should be increasing with time, providing additional fuel for star formation.  
Given the dynamical timescale of the DR21 filament (1.5 Myr) and its density (about 700 cm$^{-3}$) \citep{Schneider2010}, we estimate that the infall rate of material is a few tens of solar masses per Myr, which will not cause a substantial change to the gas mass or gas density over the $\sim$ 10 Myr of our simulations.
Along the same lines, our model gas distribution is not as clumpy and complicated as the distribution of gas in likely to be in real star forming regions. 
Additional localized star formation may occur as higher density regions of gas collapse further.

The MYStIX observations \citep{Getman2014}, and others \citep[e.g.][]{Beccari2017} suggest that there is an age spread or gradient of the stars in many of these young clusters.
Their observations suggest that in most clusters, the younger stars are found closer to the cluster centre.
While our models do not include continuous star formation that could produce such a gradient, we can comment on the timescale over which such a gradient could be preserved.
If we tag the stars by their location at the start of our simulation, we see that the stars in the outskirts remain primarily in the outskirts for about 2~Myr in the DR21 simulations, but after that time, they become mixed in with the stars from the inner regions.
This mixing should be affected by the boundedness of the stars to their birth complex, and may also depend on the duration and spatial extent of the star-forming event.
Open clusters do not show an age gradient, perhaps because of this dynamical mixing.
A more careful study of the age distribution of stars in embedded clusters, and the subsequent dynamical evolution of the population, could be very useful for understanding how and where stars form, and what subsequent observational signatures of that star formation process may exist.

An interesting question in the study of young clusters concerns mass segregation.
Are massive stars born at the centres of clusters (``primordial mass segregation''), or do young clusters become mass segregated due to dynamical processes? 
In our simulations, we impose a random distribution of stellar masses on our initial conditions, so we cannot comment in detail on the dynamical evolution of a variety of initial mass distributions.
A recent paper \citep{Dominguez2017} may point to an answer, however.
Those authors used detailed N-body simulations of a variety of initial cluster configurations and initial primordial mass segregation, or lack thereof.
They also included a non-evolving gas potential.
In all cases, their clusters showed evidence of mass segregation by the time the system was spherical, no matter what initial conditions they chose. These results are similar to simulations with clumpy initial substructure but no gas \citep[e.g.][]{McMillan2007, Parker+2016}
A preliminary analysis of our simulations suggests that we see the same effect, although as we follow the simulations further in time, it appears that the mass segregation becomes more pronounced.
A more sophisticated analysis of simulations with different distributions of stellar masses would be interesting, particularly to check if interactions with the gas could modify this dynamical process.

We have neglected binary stars in our simulations, even though we expect that the binary fraction in young clusters should be quite high.
The likely effect would be to increase the stellar mass of the system, since we have assumed that each observed star is a singleton, but in fact each point of light has a significant chance of being a double or even a higher-order multiple.
There could also be increased dynamical interaction between the stars, especially as the stellar density increases.
Binaries may also act as an internal energy source, preventing the density from continuing to increase in real clusters.
Future dynamical simulations are being planned with a realistic binary fraction and distribution of the initial binary properties to test these suggestions.

\section*{Acknowledgements}

A.S. is funded by the Natural Sciences and Engineering Research Council of Canada. S.J. gratefully acknowledges the support of a postgraduate scholarship from the School of Physics \& Astronomy at Cardiff University and the UK Science and Technology Facilities Council.
The simulations presented here use Numpy \citep{VanDerWalt2011} and Matplotlib \citep{Hunter2007}. 
The authors thank Inti Pelupessy for his help with the FRESCO software used to create the synthetic observations.
We are grateful to the referee, Richard Parker, whose comments helped improve the article.
\bibliographystyle{mnras}
\bibliography{SubclusterEvolution}\label{lastpage}

\begin{thebibliography}{}
\makeatletter
\relax
\def\mn@urlcharsother{\let\do\@makeother \do\$\do\&\do\#\do\^\do\_\do\%\do\~}
\def\mn@doi{\begingroup\mn@urlcharsother \@ifnextchar [ {\mn@doi@}
  {\mn@doi@[]}}
\def\mn@doi@[#1]#2{\def\@tempa{#1}\ifx\@tempa\@empty \href
  {http://dx.doi.org/#2} {doi:#2}\else \href {http://dx.doi.org/#2} {#1}\fi
  \endgroup}
\def\mn@eprint#1#2{\mn@eprint@#1:#2::\@nil}
\def\mn@eprint@arXiv#1{\href {http://arxiv.org/abs/#1} {{\tt arXiv:#1}}}
\def\mn@eprint@dblp#1{\href {http://dblp.uni-trier.de/rec/bibtex/#1.xml}
  {dblp:#1}}
\def\mn@eprint@#1:#2:#3:#4\@nil{\def\@tempa {#1}\def\@tempb {#2}\def\@tempc
  {#3}\ifx \@tempc \@empty \let \@tempc \@tempb \let \@tempb \@tempa \fi \ifx
  \@tempb \@empty \def\@tempb {arXiv}\fi \@ifundefined
  {mn@eprint@\@tempb}{\@tempb:\@tempc}{\expandafter \expandafter \csname
  mn@eprint@\@tempb\endcsname \expandafter{\@tempc}}}

\bibitem[\protect\citeauthoryear{{Banerjee} \& {Kroupa}}{{Banerjee} \&
  {Kroupa}}{2015}]{Banerjee2015}
{Banerjee} S.,  {Kroupa} P.,  2015, \mn@doi [\mnras] {10.1093/mnras/stu2445},
  \href {http://adsabs.harvard.edu/abs/2015MNRAS.447..728B} {447, 728}

\bibitem[\protect\citeauthoryear{{Beccari} et~al.,}{{Beccari}
  et~al.}{2017}]{Beccari2017}
{Beccari} G.,  et~al., 2017, preprint, \href
  {http://adsabs.harvard.edu/abs/2017arXiv170509496B} {} (\mn@eprint {arXiv}
  {1705.09496})

\bibitem[\protect\citeauthoryear{{Benincasa}, {Wadsley}, {Couchman}  \&
  {Keller}}{{Benincasa} et~al.}{2016}]{Benincasa2016}
{Benincasa} S.~M.,  {Wadsley} J.,  {Couchman} H.~M.~P.,   {Keller} B.~W.,
  2016, \mn@doi [\mnras] {10.1093/mnras/stw1741}, \href
  {http://adsabs.harvard.edu/abs/2016MNRAS.462.3053B} {462, 3053}

\bibitem[\protect\citeauthoryear{{Cartwright} \& {Whitworth}}{{Cartwright} \&
  {Whitworth}}{2004}]{Cartwright+2004}
{Cartwright} A.,  {Whitworth} A.~P.,  2004, \mn@doi [\mnras]
  {10.1111/j.1365-2966.2004.07360.x}, \href
  {http://adsabs.harvard.edu/abs/2004MNRAS.348..589C} {348, 589}

\bibitem[\protect\citeauthoryear{{Dom{\'{\i}}nguez}, {Fellhauer}, {Bla{\~n}a},
  {Farias}  \& {Dabringhausen}}{{Dom{\'{\i}}nguez}
  et~al.}{2017}]{Dominguez2017}
{Dom{\'{\i}}nguez} R.,  {Fellhauer} M.,  {Bla{\~n}a} M.,  {Farias} J.~P.,
  {Dabringhausen} J.,  2017, \mn@doi [\mnras] {10.1093/mnras/stx1883}, \href
  {http://adsabs.harvard.edu/abs/2017MNRAS.472..465D} {472, 465}

\bibitem[\protect\citeauthoryear{{Feigelson} et~al.,}{{Feigelson}
  et~al.}{2013}]{MYSTIXOverview2013}
{Feigelson} E.~D.,  et~al., 2013, \mn@doi [\apjs] {10.1088/0067-0049/209/2/26},
  \href {http://adsabs.harvard.edu/abs/2013ApJS..209...26F} {209, 26}

\bibitem[\protect\citeauthoryear{{Fellhauer}, {Wilkinson}  \&
  {Kroupa}}{{Fellhauer} et~al.}{2009}]{Fellhauer2009}
{Fellhauer} M.,  {Wilkinson} M.~I.,   {Kroupa} P.,  2009, \mn@doi [\mnras]
  {10.1111/j.1365-2966.2009.15009.x}, \href
  {http://adsabs.harvard.edu/abs/2009MNRAS.397..954F} {397, 954}

\bibitem[\protect\citeauthoryear{{Foster} et~al.,}{{Foster}
  et~al.}{2015}]{Foster2015}
{Foster} J.~B.,  et~al., 2015, \mn@doi [\apj] {10.1088/0004-637X/799/2/136},
  \href {http://adsabs.harvard.edu/abs/2015ApJ...799..136F} {799, 136}

\bibitem[\protect\citeauthoryear{{Fujii}, {Iwasawa}, {Funato}  \&
  {Makino}}{{Fujii} et~al.}{2007}]{Fujii2007}
{Fujii} M.,  {Iwasawa} M.,  {Funato} Y.,   {Makino} J.,  2007, \mn@doi [\pasj]
  {10.1093/pasj/59.6.1095}, \href
  {http://adsabs.harvard.edu/abs/2007PASJ...59.1095F} {59, 1095}

\bibitem[\protect\citeauthoryear{{Fujii}, {Saitoh}  \& {Portegies
  Zwart}}{{Fujii} et~al.}{2012}]{Fujii2012}
{Fujii} M.~S.,  {Saitoh} T.~R.,   {Portegies Zwart} S.~F.,  2012, \mn@doi
  [\apj] {10.1088/0004-637X/753/1/85}, \href
  {http://adsabs.harvard.edu/abs/2012ApJ...753...85F} {753, 85}

\bibitem[\protect\citeauthoryear{{Getman} et~al.,}{{Getman}
  et~al.}{2014}]{Getman2014}
{Getman} K.~V.,  et~al., 2014, \mn@doi [\apj] {10.1088/0004-637X/787/2/108},
  \href {http://adsabs.harvard.edu/abs/2014ApJ...787..108G} {787, 108}

\bibitem[\protect\citeauthoryear{{Goodwin} \& {Bastian}}{{Goodwin} \&
  {Bastian}}{2006}]{GoodwinBastian2006}
{Goodwin} S.~P.,  {Bastian} N.,  2006, \mn@doi [\mnras]
  {10.1111/j.1365-2966.2006.11078.x}, \href
  {http://adsabs.harvard.edu/abs/2006MNRAS.373..752G} {373, 752}

\bibitem[\protect\citeauthoryear{{Gratton}, {Carretta}  \&
  {Bragaglia}}{{Gratton} et~al.}{2012}]{Gratton2012}
{Gratton} R.~G.,  {Carretta} E.,   {Bragaglia} A.,  2012, \mn@doi [\aapr]
  {10.1007/s00159-012-0050-3}, \href
  {http://adsabs.harvard.edu/abs/2012A%26ARv..20...50G} {20, 50}

\bibitem[\protect\citeauthoryear{{Hubber}, {Allison}, {Smith}  \&
  {Goodwin}}{{Hubber} et~al.}{2013}]{Hubber2013}
{Hubber} D.~A.,  {Allison} R.~J.,  {Smith} R.,   {Goodwin} S.~P.,  2013,
  \mn@doi [\mnras] {10.1093/mnras/sts694}, \href
  {http://adsabs.harvard.edu/abs/2013MNRAS.430.1599H} {430, 1599}

\bibitem[\protect\citeauthoryear{{Hunter}}{{Hunter}}{2007}]{Hunter2007}
{Hunter} J.~D.,  2007, \mn@doi [Computing in Science and Engineering]
  {10.1109/MCSE.2007.55}, \href
  {http://adsabs.harvard.edu/abs/2007CSE.....9...90H} {9, 90}

\bibitem[\protect\citeauthoryear{{Jaffa}, {Whitworth}  \& {Lomax}}{{Jaffa}
  et~al.}{2017}]{Jaffa2017}
{Jaffa} S.~E.,  {Whitworth} A.~P.,   {Lomax} O.,  2017, \mn@doi [\mnras]
  {10.1093/mnras/stw3140}, \href
  {http://adsabs.harvard.edu/abs/2017MNRAS.466.1082J} {466, 1082}

\bibitem[\protect\citeauthoryear{{Kroupa}}{{Kroupa}}{2001}]{Kroupa2001}
{Kroupa} P.,  2001, \mn@doi [\mnras] {10.1046/j.1365-8711.2001.04022.x}, \href
  {http://adsabs.harvard.edu/abs/2001MNRAS.322..231K} {322, 231}

\bibitem[\protect\citeauthoryear{{Kruijssen}, {Pelupessy}, {Lamers}, {Portegies
  Zwart}  \& {Icke}}{{Kruijssen} et~al.}{2011}]{Kruijssen2011}
{Kruijssen} J.~M.~D.,  {Pelupessy} F.~I.,  {Lamers} H.~J.~G.~L.~M.,  {Portegies
  Zwart} S.~F.,   {Icke} V.,  2011, \mn@doi [\mnras]
  {10.1111/j.1365-2966.2011.18467.x}, \href
  {http://adsabs.harvard.edu/abs/2011MNRAS.414.1339K} {414, 1339}

\bibitem[\protect\citeauthoryear{{Kuhn} et~al.,}{{Kuhn}
  et~al.}{2014}]{Kuhn2014}
{Kuhn} M.~A.,  et~al., 2014, \mn@doi [\apj] {10.1088/0004-637X/787/2/107},
  \href {http://adsabs.harvard.edu/abs/2014ApJ...787..107K} {787, 107}

\bibitem[\protect\citeauthoryear{{Kuhn}, {Getman}  \& {Feigelson}}{{Kuhn}
  et~al.}{2015}]{Kuhn2015}
{Kuhn} M.~A.,  {Getman} K.~V.,   {Feigelson} E.~D.,  2015, \mn@doi [\apj]
  {10.1088/0004-637X/802/1/60}, \href
  {http://adsabs.harvard.edu/abs/2015ApJ...802...60K} {802, 60}

\bibitem[\protect\citeauthoryear{{Lada} \& {Lada}}{{Lada} \&
  {Lada}}{2003}]{LadaLada2003}
{Lada} C.~J.,  {Lada} E.~A.,  2003, \mn@doi [\araa]
  {10.1146/annurev.astro.41.011802.094844}, \href
  {http://adsabs.harvard.edu/abs/2003ARA%26A..41...57L} {41, 57}

\bibitem[\protect\citeauthoryear{{Massi}, {Giannetti}, {Di Carlo}, {Brand},
  {Beltr{\'a}n}  \& {Marconi}}{{Massi} et~al.}{2015}]{Massi2015}
{Massi} F.,  {Giannetti} A.,  {Di Carlo} E.,  {Brand} J.,  {Beltr{\'a}n} M.~T.,
    {Marconi} G.,  2015, \mn@doi [\aap] {10.1051/0004-6361/201424388}, \href
  {http://adsabs.harvard.edu/abs/2015A%26A...573A..95M} {573, A95}

\bibitem[\protect\citeauthoryear{{McMillan}, {Vesperini}  \& {Portegies
  Zwart}}{{McMillan} et~al.}{2007}]{McMillan2007}
{McMillan} S.~L.~W.,  {Vesperini} E.,   {Portegies Zwart} S.~F.,  2007, \mn@doi
  [\apjl] {10.1086/511763}, \href
  {http://adsabs.harvard.edu/abs/2007ApJ...655L..45M} {655, L45}

\bibitem[\protect\citeauthoryear{{Myers}}{{Myers}}{2010}]{Myers2010}
{Myers} P.~C.,  2010, \mn@doi [\apj] {10.1088/0004-637X/714/2/1280}, \href
  {http://adsabs.harvard.edu/abs/2010ApJ...714.1280M} {714, 1280}

\bibitem[\protect\citeauthoryear{{Offner}, {Hansen}  \& {Krumholz}}{{Offner}
  et~al.}{2009}]{Offner2009}
{Offner} S.~S.~R.,  {Hansen} C.~E.,   {Krumholz} M.~R.,  2009, \mn@doi [\apjl]
  {10.1088/0004-637X/704/2/L124}, \href
  {http://adsabs.harvard.edu/abs/2009ApJ...704L.124O} {704, L124}

\bibitem[\protect\citeauthoryear{{Parker} \& {Dale}}{{Parker} \&
  {Dale}}{2013}]{Parker2013}
{Parker} R.~J.,  {Dale} J.~E.,  2013, \mn@doi [\mnras] {10.1093/mnras/stt517},
  \href {http://adsabs.harvard.edu/abs/2013MNRAS.432..986P} {432, 986}

\bibitem[\protect\citeauthoryear{{Parker} \& {Wright}}{{Parker} \&
  {Wright}}{2016}]{Parker2016}
{Parker} R.~J.,  {Wright} N.~J.,  2016, \mn@doi [\mnras]
  {10.1093/mnras/stw087}, \href
  {http://adsabs.harvard.edu/abs/2016MNRAS.457.3430P} {457, 3430}

\bibitem[\protect\citeauthoryear{{Parker}, {Wright}, {Goodwin}  \&
  {Meyer}}{{Parker} et~al.}{2014}]{Parker2014}
{Parker} R.~J.,  {Wright} N.~J.,  {Goodwin} S.~P.,   {Meyer} M.~R.,  2014,
  \mn@doi [\mnras] {10.1093/mnras/stt2231}, \href
  {http://adsabs.harvard.edu/abs/2014MNRAS.438..620P} {438, 620}

\bibitem[\protect\citeauthoryear{{Parker}, {Goodwin}, {Wright}, {Meyer}  \&
  {Quanz}}{{Parker} et~al.}{2016}]{Parker+2016}
{Parker} R.~J.,  {Goodwin} S.~P.,  {Wright} N.~J.,  {Meyer} M.~R.,   {Quanz}
  S.~P.,  2016, \mn@doi [\mnras] {10.1093/mnrasl/slw061}, \href
  {http://adsabs.harvard.edu/abs/2016MNRAS.459L.119P} {459, L119}

\bibitem[\protect\citeauthoryear{{Pelupessy} \& {Portegies Zwart}}{{Pelupessy}
  \& {Portegies Zwart}}{2012}]{Pelupessy2012}
{Pelupessy} F.~I.,  {Portegies Zwart} S.,  2012, \mn@doi [\mnras]
  {10.1111/j.1365-2966.2011.20137.x}, \href
  {http://adsabs.harvard.edu/abs/2012MNRAS.420.1503P} {420, 1503}

\bibitem[\protect\citeauthoryear{{Pelupessy}, {van Elteren}, {de Vries},
  {McMillan}, {Drost}  \& {Portegies Zwart}}{{Pelupessy}
  et~al.}{2013}]{AMUSE2013}
{Pelupessy} F.~I.,  {van Elteren} A.,  {de Vries} N.,  {McMillan} S.~L.~W.,
  {Drost} N.,   {Portegies Zwart} S.~F.,  2013, \mn@doi [\aap]
  {10.1051/0004-6361/201321252}, \href
  {http://adsabs.harvard.edu/abs/2013A%26A...557A..84P} {557, A84}

\bibitem[\protect\citeauthoryear{{Plummer}}{{Plummer}}{1911}]{Plummer1911}
{Plummer} H.~C.,  1911, \mn@doi [\mnras] {10.1093/mnras/71.5.460}, \href
  {http://adsabs.harvard.edu/abs/1911MNRAS..71..460P} {71, 460}

\bibitem[\protect\citeauthoryear{{Portegies Zwart} et~al.,}{{Portegies Zwart}
  et~al.}{2009}]{AMUSE2009}
{Portegies Zwart} S.,  et~al., 2009, \mn@doi [\na]
  {10.1016/j.newast.2008.10.006}, \href
  {http://adsabs.harvard.edu/abs/2009NewA...14..369P} {14, 369}

\bibitem[\protect\citeauthoryear{{Proszkow} \& {Adams}}{{Proszkow} \&
  {Adams}}{2009}]{Proszkow2009}
{Proszkow} E.-M.,  {Adams} F.~C.,  2009, \mn@doi [\apjs]
  {10.1088/0067-0049/185/2/486}, \href
  {http://adsabs.harvard.edu/abs/2009ApJS..185..486P} {185, 486}

\bibitem[\protect\citeauthoryear{{Reid} \& {Wilson}}{{Reid} \&
  {Wilson}}{2006}]{Reid2006}
{Reid} M.~A.,  {Wilson} C.~D.,  2006, \mn@doi [\apj] {10.1086/503824}, \href
  {http://adsabs.harvard.edu/abs/2006ApJ...644..990R} {644, 990}

\bibitem[\protect\citeauthoryear{{Schneider}, {Csengeri}, {Bontemps}, {Motte},
  {Simon}, {Hennebelle}, {Federrath}  \& {Klessen}}{{Schneider}
  et~al.}{2010}]{Schneider2010}
{Schneider} N.,  {Csengeri} T.,  {Bontemps} S.,  {Motte} F.,  {Simon} R.,
  {Hennebelle} P.,  {Federrath} C.,   {Klessen} R.,  2010, \mn@doi [\aap]
  {10.1051/0004-6361/201014481}, \href
  {http://adsabs.harvard.edu/abs/2010A%26A...520A..49S} {520, A49}

\bibitem[\protect\citeauthoryear{{Springel}}{{Springel}}{2005}]{Springel2005}
{Springel} V.,  2005, \mn@doi [\mnras] {10.1111/j.1365-2966.2005.09655.x},
  \href {http://adsabs.harvard.edu/abs/2005MNRAS.364.1105S} {364, 1105}

\bibitem[\protect\citeauthoryear{{Stutz}}{{Stutz}}{2017}]{Stutz2017}
{Stutz} A.~M.,  2017, preprint, \href
  {http://adsabs.harvard.edu/abs/2017arXiv170505838S} {} (\mn@eprint {arXiv}
  {1705.05838})

\bibitem[\protect\citeauthoryear{{Van Der Walt}, {Colbert}  \&
  {Varoquaux}}{{Van Der Walt} et~al.}{2011}]{VanDerWalt2011}
{Van Der Walt} S.,  {Colbert} S.~C.,   {Varoquaux} G.,  2011, preprint, \href
  {http://adsabs.harvard.edu/abs/2011arXiv1102.1523V} {} (\mn@eprint {arXiv}
  {1102.1523})

\makeatother
\end{thebibliography}
\end{document}